\documentclass[11pt,twoside]{article}
\usepackage{geometry}
\usepackage{graphicx}
\usepackage[english]{babel}
\pdfoutput=1
\usepackage[utf8]{inputenc}
\usepackage{interval}
\usepackage{comment}
\usepackage{amssymb}  
\usepackage{nccmath}  
\usepackage{natbib}
\usepackage{enumitem}
\usepackage{multirow}
\usepackage{float}  
\usepackage{cancel}
\usepackage{algorithm}
\usepackage[table]{xcolor}
\usepackage{longtable}
\usepackage{pgf, tikz}
\usepackage{tikz-network}
\usetikzlibrary{positioning, shapes, shadows, arrows, patterns, automata}
\usepackage{pgfplots}
\pgfplotsset{compat=newest}

\tikzset{node_style/.style={draw,circle,line width=.1mm, inner sep=0,font=\fontsize{7}{10}\selectfont}}
\tikzset{edge_style/.style={draw=black, ultra thick, line width=.1mm,font=\fontsize{7}{10}\selectfont}}
\usepackage{xfrac}
\usepackage{array,multirow}

\usepackage[flushleft]{threeparttable}

\makeatletter
\newcommand{\thickhline}{%
    \noalign {\ifnum 0=`}\fi \hrule height 1pt
    \futurelet \reserved@a \@xhline
}
\newcolumntype{"}{@{\hskip\tabcolsep\vrule width 1pt\hskip\tabcolsep}}
\makeatother

\newcolumntype{P}[1]{>{\centering\arraybackslash}p{#1}}
\newcolumntype{M}[1]{>{\centering\arraybackslash}m{#1}}

\newcolumntype{?}{!{\vrule width 1pt}}

\usepackage{makecell}

\usepackage{esvect}

\usepackage{listings}
\newcommand{\listingsfont}{\ttfamily}

\usepackage{hhline}  

\usepackage{subcaption}
\usepackage{adjustbox}

\newcommand{\tikzmark}[1]{\tikz[overlay,remember picture] \node (#1) {};}

\newcolumntype{C}[1]{>{\centering\arraybackslash}m{#1}}

\usepackage[customcolors]{hf-tikz}
\hfsetfillcolor{gray!10}
\hfsetbordercolor{gray}

\usepackage{hyperref}
\usepackage{amsfonts}
\usepackage{interval}
\usepackage{esvect}
\usepackage{comment}
\usepackage{amsthm,amsmath,amssymb,amsfonts,nccmath,mathtools,textcomp,graphicx,makeidx,color,gensymb,mhchem,url}
\usepackage{natbib,cite}
\usepackage{indentfirst}
\usepackage{enumitem}
\usepackage{multirow}
\usepackage{float}
\usepackage{cancel}
\usepackage{algorithm}
\usepackage[table]{xcolor}
\usepackage{longtable}
\usepackage{pgf, tikz}
\usepackage{tikz-network}
\usetikzlibrary{positioning, shapes, shadows, arrows, snakes,patterns}
\usetikzlibrary{arrows, automata}
\usepackage{pgfplots}

\tikzset{node_style/.style={draw,circle,line width=.1mm, inner sep=0,font=\fontsize{7}{10}\selectfont}}
\tikzset{edge_style/.style={draw=black, ultra thick, line width=.1mm,font=\fontsize{7}{10}\selectfont}}

\usepackage{array,multirow}

\usepackage{array}
\newcolumntype{C}[1]{>{\centering\arraybackslash}m{#1}}

\usepackage[flushleft]{threeparttable}

\usepackage{pdflscape}

\newcolumntype{"}{@{\hskip\tabcolsep\vrule width 1pt\hskip\tabcolsep}}
\makeatother

\newcolumntype{P}[1]{>{\centering\arraybackslash}p{#1}}
\newcolumntype{M}[1]{>{\centering\arraybackslash}m{#1}}

\usepackage{array}
\newcolumntype{?}{!{\vrule width 1pt}}

\usepackage{makecell}

\usepackage{esvect}

\usepackage{listings}

\usepackage{hhline}  
\usepackage[table]{xcolor}

\def\titleab#1{{\Large\bf  \begin{flushleft} #1 \vspace{0pt} \end{flushleft}}}
\def\authors#1{{ \bf \begin{flushleft} #1 \vspace{0pt} \end{flushleft}}}
\def\university#1{{ \begin{flushleft} #1 \vspace{0pt} \end{flushleft}}}
\def\inst#1{\unskip$^{#1}$}

\newcommand{\keywords}[1]{\bigskip \noindent {\bf Keywords:} \ #1}

\newtheorem{definition}{Definition}[section]
\newtheorem{example}{Example}[section]
\newtheorem{note}{Note}[section]

\usepackage[font=scriptsize]{caption} 
\usepackage[font=scriptsize]{subcaption}

\usepackage{adjustbox}

\usepackage{array}
\newcolumntype{C}[1]{>{\centering\arraybackslash}m{#1}}

\usepackage[flushleft]{threeparttable}
\usepackage[customcolors]{hf-tikz}
\hfsetfillcolor{gray!10}
\hfsetbordercolor{gray}

\newcommand\blfootnote[1]{%
  \begingroup
  \renewcommand\thefootnote{}\footnote{#1}%
  \addtocounter{footnote}{-1}%
  \endgroup
}

\begin{document}

\titleab{Community Detection in Interval-Weighted Networks}


\authors{
  H\'elder Alves\textsuperscript{*}\inst{1}\blfootnote{* Corresponding author: H\'elder Alves, helder.alves@isssp.pt, Porto, Portugal}
  Paula Brito\inst{2},
  Pedro Campos\inst{2}
}

\university{
  \inst{1} ISSSP, Porto Institute of Social Work \& LIAAD INESC TEC, Portugal, \href{mailto:author1@mail.pt}{helder.alves@isssp.pt}\\
  \inst{2} FEP, University of Porto \& LIAAD INESC TEC, Portugal, \href{mailto:author2@mail.pt}{mpbrito@fep.up.pt}\\
}

\begin{abstract}
In this paper we introduce and develop the concept of Interval-Weighted Networks (IWN), a novel approach in Social Network Analysis, where the edge weights are represented by closed intervals composed with precise information, comprehending intrinsic variability. We extend IWN for both Newman's modularity and modularity gain and the Louvain algorithm (LA), considering a tabular representation of networks by contingency tables. We apply our methodology in a real-world commuter network in mainland Portugal between the twenty three NUTS 3 regions. The optimal partition of regions is developed and compared using two new different approaches, designated as ``Classic Louvain'' (CL) and ``Hybrid Louvain'' (HL), which allow taking into account the variability observed in the original network, thereby minimizing the loss of information present in the raw data. Our findings suggest the division of the twenty three Portuguese regions in three main communities. However, we find different geographical partitions according to the community detection methodology used. This analysis can be useful in many real-world applications, since it takes into account that the weights may vary within the ranges, rather than being constant.\\

\keywords{Community Detection, Interval-Weighted Networks, weighted networks, Commuter networks, Louvain algorithm}

\end{abstract}


\section{Introduction}
\label{intro}
Nowadays, we are increasingly living in a complex and interconnected world, where the amount of data available as well as the technology required to have access to and mine/explore this data (computational capacity) has become increasingly affordable. As a consequence, on-line networking services like Facebook, Twitter, WhatsApp, Instagram, among others, registered an astounding growth reaching hundreds of millions of users. Regardless of the context and size of these networks, in classical graph theory, they are usually represented in the form of binary or weighted networks, where the weights on the edges are assumed to be constant~\citep{Newman:2004jh}.
However, in real-world applications these weights may vary within ranges rather than being constant~\citep{Hu:2008vt}. To better model such variability of weights in a network, instead of using constants (real numbers) and associated methods to represent the information present in the edges, we represent weights as intervals. A representation of these values in the form of closed intervals composed with precise information, can be more meaningful and useful in a dynamic environment than a point-valued output, as these intervals contain more information in expressing raw data variability, thereby minimizing the loss of information~\citep{NoirhommeFraiture:2011ck,Couso:2014du,Grzegorzewski:2016fq}. Taking into account the variability of edge weights in the form of closed intervals, we call our networks \textit{interval-weighted networks} (IWN). Figure~\ref{chp3_fig:IWN} below shows an example of an undirected interval-weighted network.

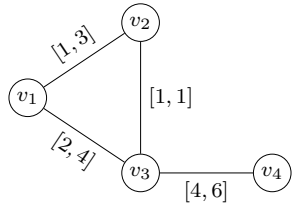
\begin{figure}[ht]
 \centering      
	\begin{tikzpicture}[inner sep=0pt, minimum size=5mm, auto,
   	node_style/.style={draw,circle,line width=.1mm, font=\fontsize{8}{10}\selectfont},
   	edge_style/.style={draw=black, ultra thick, line width=.1mm}]
    \node[node_style] (v1) at (0,1) {$v_1$};
    \node[node_style] (v2) at (1.5,2) {$v_2$};
    \node[node_style] (v3) at (1.5,0) {$v_3$};
    \node[node_style] (v4) at (3.25,0) {$v_4$};
    \draw[edge_style]  (v1) edge node[above,sloped,pos=0.5,font=\fontsize{8}{10}\selectfont] {$[1,3]$} (v2);
    \draw[edge_style]  (v1) edge node[below,sloped,pos=0.5,font=\fontsize{8}{10}\selectfont] {$[2,4]$} (v3);
    \draw[edge_style]  (v2) edge node[right=0.1,pos=0.5,font=\fontsize{8}{10}\selectfont] {$[1,1]$} (v3);
    \draw[edge_style]  (v3) edge node[below,pos=0.5,font=\fontsize{8}{10}\selectfont] {$[4,6]$} (v4);
	\end{tikzpicture}
    \caption{Example of an undirected Interval-Weighted Network (IWN).} \label{chp3_fig:IWN}
\end{figure}    	

One of the most important/studied features in networks is the existence of a community structure. Identifying these communities (or clusters), which are tightly (densely) connected internally, and less with the rest of the network, is helpful to a better understanding and visualisation of the whole network~\citep{Wasserman:1994wu,Girvan:2001ez,Newman:2003wd,Guimera:2003el,BOCCALETTI:2006gb,Farkas:2007bl}. In order to derive a measure of quality of a partition, even without such prior information, \citet{Newman:2004te} introduced a \textit{quality function} known as \textit{modularity} $(Q)$, which is a quantitative criterion to evaluate the quality of a certain partition. Roughly speaking, Newman and Girvan's modularity compares a given network to a network with the same degree distribution of ties over the nodes placed at random. To optimize the Girvan-Newman modularity, i.e., find a global maximum for $Q$, one of the fastest and best methods in terms of efficiency and accuracy is the Louvain algorithm (LA)~\citep{Lancichinetti:2009bb}. LA is a greedy hierarchical clustering algorithm introduced in 2008 by V. Blondel, J-L. Guillaume, R. Lambiotte~\citep{Blondel:2008do}, and aims at partitioning a network into non-overlapping communities by heuristically optimizing the Girvan-Newman modularity.\\
Our goal in this paper is twofold, as we aim at extending to IWN both: (i) Newman's modularity and modularity gain for weighted networks and (ii) the Louvain algorithm (LA), considering a tabular representation of networks (by contingency tables)~\citep{Traag:2014tj}. Finally, we apply our methodology in a real-world network, to put in evidence the community structure that emerges from the movements of daily commuters in mainland Portugal, between the twenty three NUTS 3 regions.\\
This paper is organized as follows. In the next section (Section~\ref{Interval_Analysis}), we introduce the basic terms and concepts of interval arithmetic and interval order relations, and based on our purpose to capture the maximum variability of an interval, a new approach for ranking intervals is proposed. The following Section~\ref{CD_WN} begins with the extension of modularity and modularity gain considering a tabular representation of networks (by contingency tables). In Section~\ref{CD_IWN} we generalize these notions to the case of interval-weighted networks (IWN), first defining new measures to evaluate the difference between two intervals, then extending the modularity, modularity gain and the LA to deal with IWN, developing a methodology based on two major methods: ``Classic Louvain'' (CL)  and ``Hybrid Louvain'' (HL). In Section~\ref{example_commuters}, we summarize and discuss the results of applying our community detection methodology in a Portuguese commuters network, between the twenty three NUTS 3 regions. Finally, Section~\ref{conclusion} presents the conclusions of our study and proposes some directions for future work.
\newpage

\section{Interval Analysis}
\label{Interval_Analysis}

Interval analysis is a methodology based on an arithmetic defined on sets of real intervals, rather than sets of real numbers. An interval operation produces two values, i.e., \textit{lower} and \textit{upper} endpoints of the resulting interval such that the true result certainly lies between those points, and the ``accuracy'' of the result is evaluated by the width of the interval~\citep{Moore:1979ug,Moore:2009uc,Dawood:2011vh}. In the vast majority of existing literature concerning interval analysis, intervals are considered as disjunctive sets representing incomplete information \textit{(epistemic view)}. However, our approach of using intervals is that a closed interval may be used to model the precise information of an objective entity that comprehends intrinsic variability \textit{(ontic view)}, i.e., an interval $A$ is a value of a set-valued variable $X$, so we can write $X=A$. Such intervals are called conjunctive and may, for example, represent ranges of fluctuations of some measurements, time interval spanned by an activity, among others~\citep{Couso:2014du,Grzegorzewski:2016fq}.

\subsection{Classic interval arithmetic and its pitfalls}
\label{Subsection_2.1}

Let $\underline{x},\overline{x} \in \mathbb{R}$ such that $\underline{x} \leqslant \overline{x}$. An interval number $[\underline{x},\overline{x}]$ is a closed bounded nonempty real interval, given by
$
[\underline{x},\overline{x}]=\{x \in \mathbb{R}\colon \underline{x} \leqslant x \leqslant \overline{x}\}
$,
where $\underline{x}=\min([\underline{x},\overline{x}])$ and $\overline{x}=\max([\underline{x},\overline{x}])$ are called, respectively, the \textup{lower} and \textup{upper} bounds (endpoints) of $[\underline{x},\overline{x}]$.
The set $[\mathbb{R}]$ of interval numbers is a subset of the powerset of $\mathbb{R}$ such that 
$
[\mathbb{R}]=\big\{X \in \wp\, (\mathbb{R})\colon (\exists \underline{x} \in \mathbb{R})\, (\exists \overline{x} \in \mathbb{R})\, (X=[\underline{x},\overline{x}])\big\}.
$
Since, corresponding to each pair of real constants $\underline{x},\overline{x}\ (\underline{x}\leqslant\overline{x})$ there exists a closed interval $[\underline{x},\overline{x}]$, the set of interval numbers is \textit{infinite}. We say that $X$ is \textup{degenerate} if $\underline{x}=\overline{x}$. By convention, a degenerate interval $[x,x]$ is identified with the real number $x$ (e.g. $1=[1,1]$).
For any two intervals $X=[\underline{x}, \overline{x}]$ and $Y=[\underline{y}, \overline{y}]$, in terms of the intervals' endpoints, the four classical operations of real arithmetic can be extended to intervals as follows~\citep{Moore:1979ug}:
\begin{itemize}
\setlength\itemsep{0.5pt} 
\item 
Interval addition, 
$
X+Y=[\underline{x}, \overline{x}] + [\underline{y}, \overline{y}] = [\underline{x}+\underline{y}, \overline{x}+\overline{y}]
$;
\item Interval multiplication,
$
X\cdot Y=[\underline{x}, \overline{x}] \cdot [\underline{y}, \overline{y}]=\big\lbrack\min\{\underline{x}\underline{y},\underline{x}\overline{y},\overline{x}\underline{y},\overline{x}\overline{y}\}, \max\{\underline{x}\underline{y},\underline{x}\overline{y},\overline{x}\underline{y},\overline{x}\overline{y}\}\big\rbrack
$;
\item Interval subtraction,
$
X-Y=X+(-Y)
$
where $-Y=[-\overline y,-\underline y]$ (reversal of endpoints)\footnote{It should be noted that the subtraction of two equal intervals is not $\interval{0}{0}$ (except for degenerate intervals). This is because $X-X=\{x-y\colon x \in X,\, y \in X \}$, rather than $\{x-x\colon x \in X\}$~\citep{Jaulin:2001tq}. For example, $\interval{1}{2}-\interval{1}{2}=\interval{-1}{1}$.}.
\item 
Interval division for any $X\in\mathbb{R}$ and any $Y\in\lbrack\mathbb{R}\rbrack_{\widetilde{0}}$, is defined by
$
X \div Y = X\cdot(Y^{-1})
$,
where $Y^{-1}=1/Y=[1/ \overline y,1/ \underline y]$, assuming that $0 \not\in Y$.
\end{itemize}

Intervals can also be represented by their \textit{midpoint (or mean, or center)} $m$ and \textit{half-width} (or radius), $rad$. So, $X=\interval{\underline{x}}{\overline{x}}=\langle m(X),rad(X) \rangle$, where $m(X)=\mfrac{(\underline{x} + \overline{x})}{2}$ and $rad(X)=\mfrac{(\overline{x} - \underline{x})}{2}$. An operation whose operands are intervals $\interval{\underline{x}}{\overline{x}}$, and whose result is a point interval (or a real number) is called a \textit{point interval operation}, such as: \textit{infimum} $\inf(\interval{\underline{x}}{\overline{x}})=\min(\interval{\underline{x}}{\overline{x}})=\underline{x}$ and \textit{supremum} $\sup(\interval{\underline{x}}{\overline{x}})=\max(\interval{\underline{x}}{\overline{x}})=\overline{x}$. Therefore, the \textit{infimum} of two intervals $X=\interval{\underline{x}}{\overline{x}}$ and $Y=\interval{\underline{y}}{\overline{y}}$ is defined to be $
\inf(X,Y)=\lbrack\inf(\underline{x},\underline{y}),\inf(\overline{x},\overline{y})\rbrack$. Similarly, the \textit{supremum} of two intervals $X=\interval{\underline{x}}{\overline{x}}$ and $Y=\interval{\underline{y}}{\overline{y}}$ is defined to be $
\sup(X,Y)=\lbrack\sup(\underline{x},\underline{y}),\sup(\overline{x},\overline{y})\rbrack$~\citep{Moore:2009uc,Dawood:2011vh}. Finally, another important definition of a point interval operation is the Hausdorff distance (or metric) between two intervals~\citep{Bryant:1985wi,billard2006symbolic}:
$d(X,Y){=}d(\interval{\underline{x}}{\overline{x}}, \interval{\underline{y}}{\overline{y}}){=}\max\{|\underline{x}-\underline{y}|,|\overline{x}-\overline{y}|\}$.

However, useful properties of ordinary real arithmetic fail to hold in classical interval arithmetic. Some of the main disadvantages of the classical interval theory are~\citep{Rokne:2001kj}: (i) \textit{Interval dependency} -- subtraction and division are not the inverse operations of addition and multiplication, respectively; (ii) \textit{Distributive law does not hold} -- only a subdistributive law is valid $\left(\exists\ X,Y,Z \in [\mathbb{R}]\big)\ \big(Z \times (X+Y) \subseteq Z \times X + Z \times Y\right)$.\\
Later in this paper, in Section~\ref{CD_IWN}, Subsection~\ref{chp5_SubSec_IntervalDifferences}, these pitfalls lead us to develop new measures to assess the difference between two intervals.

\section{Community Detection in Weighted Networks based on the Contingency Table}
\label{CD_WN}

A common property of networks is their modular structure, namely their organization into modules (also called communities or clusters), in such a way that most of the links are concentrated within the modules, while there are fewer links between vertices belonging to different modules. Community detection algorithms aim at identifying the modules and, possibly, their hierarchical organization, in a graph. The modularity measure proposed by Girvan and Newman~\citep{Newman:2004te} is one of the most used and best-known functions to quantify community structure in a graph. Empirically, a high modularity value indicates a good partition. To optimize modularity, the state-of-the-art greedy method introduced by~\citet{Blondel:2008do} -- the Louvain algorithm, is generally used.

Before we approach the extension of community detection to the case of interval-weighted networks (IWN) in Section~\ref{CD_IWN}, it is important to note that our entire approach is based on the concept that the definition of modularity for a sum over vertices' pairs as~\citep{Newman:2004jh,Clauset:2004uy,2007NJPh....9..176A}, 
\begin{equation}\label{weighted_modulatity_Q}
Q^W=\mfrac{1}{2w} \sum_{i=1}^n \sum_{j=1}^n \left(w_{ij}-\mfrac{{s_i} {s_j}}{2w}\right) \delta(C_i, C_j)
\end{equation}
can be translated as the difference between \textit{the fraction of internal edges strength in the network} and \textit{the expected fraction of such edges strength placed at random while preserving the vertices strength}.

This section focuses on the generalization of modularity for weighted networks (see~(\ref{weighted_modulatity_Q})), based on a contingency table, considering the observed weights and the expected weights assuming independence between the vertices (our null model). Furthermore, we extend this approach to the Louvain algorithm (LA).

\subsection{Modularity based on the Contingency Table}
\label{chp5:sec_1}

An undirected weighted network $G^W=(V,E,W)$  with a set, $V=\{v_1,\dots,v_n\} \neq \emptyset$ of vertices, a set $E=\{e_1,\dots,e_m\}$ of edges and a set of weights or values $W=\{w_1,\dots,w_m\}$, can be represented in the form of a \textit{contingency table} (or \textit{cross-tabulation}, or \textit{crosstab})~\citep{Everitt:273919,Traag:2014tj}. Hence, based on the concept of the \textit{chi-square statistic of independence}, we can evaluate the discrepancy between \textit{the observed counts in the table} and \textit{the expected values of those counts under the null hypothesis}.\\
The generalization to the networks data type is straightforward, however, instead of ``counts'' we use ``edge weights'' of the symmetrical adjacency matrix $n\times n$, as explained below.

\begin{definition} [Contingency table for the observed weights -- $O$] \label{chp5_Def:adj_O}
Consider a contingency table of \textit{observed} weights as $O={\lbrack o_{ij}\rbrack}_{n\times n}$ with $n$ rows (``source'' vertices $i$) and $n$ columns (``destination'' vertices $j$), such that $o_{ij}=w_{ij}$ and $o_{ij}=o_{ji}>0\ (o_{ij}\in \mathbb{R^+})$, if there is an edge with weight $w_{ij}$ between vertices $(i,j)$, and zero otherwise. The \textit{marginal sums} for each row or column of the table represent the \textit{total weight} or \textit{strength} attached (or linked) to vertex $i$, denoted by $s^O_i=\sum_{j=1}^n o_{ij}$, and the total weight is $\sum_{i=1}^{n}\sum_{j=1}^{n} {o_{ij}}=2w=\sum_{i=1}^{n} s^O_i = \sum_{j=1}^{n} s^O_j$. The table of \textit{observed} weights can be written as:

\begin{equation}\label{chp5_Tab:adj_O}
\centering
\renewcommand{\arraystretch}{1.2}
\begin{tabular}{ cc| c c c c|c }
&\multicolumn{6}{ c }{\tiny To vertex} \\
 & & $v_1$ & $v_2$ & $\cdots$ & $v_n$ & $s_i^O$ \\ \cline{2-7}
\multirow{4}{*}{$O={\lbrack o_{ij}\rbrack}_{n\times n}={\rotatebox[origin=c]{90}{\tiny From vertex}}$}
 & $v_1$ & $o_{11}$ & $o_{12}$ & $\cdots$ & $o_{1n}$ & $s_1^O$ \\
 & $v_2$ & $o_{21}$ & $o_{22}$ & $\cdots$ & $o_{2n}$ & $s_2^O$ \\  	
 & $\vdots$ & $\vdots$ & $\vdots$ 	& $\ddots$ & $\vdots$ & $\vdots$ \\ 
 & $v_n$ & $o_{n1}$ & $o_{n2}$ & $\cdots$ & $o_{nn}$ & $s_n^O$ \\ \cline{2-7}
 & $s_j^O$ & $s_1^O$ & $s_2^O$ & $\cdots$ & $s_n^O$ & $2w$ \\
\end{tabular}
\end{equation}
\end{definition}

The fraction of edge weights that join vertices $i$ and $j$ is $p_{ij}=\mfrac{o_{ij}}{2w}$. Let $a_i$ and $a_j$ be the fraction of the total weight attached to vertex $i$ and vertex $j$, respectively. Then, the true values of the marginal probabilities involved are not known, they will have to be estimated, which will result in $\hat a_i=\mfrac{s_i^O}{2w}$ and $\hat a_j=\mfrac{s_j^O}{2w}$. Therefore, the \textit{expected} weights of the table, can be defined as follows:

\begin{definition} [Contingency table for the expected weights -- $E$] \label{chp5_Def:adj_E}
Let $e_{ij}$ be the expected weights assuming independence, where $e_{ij}$ is the weight that would be obtained if the hypothesis of row-column independence were true, we have
\begin{align}
e_{ij}&=2w\ \hat{a_i}\ \hat{a_j}=2w\ \frac{s_i^O}{2w}\ \frac{s_j^O}{2w} \nonumber \\
		&=\frac{s_i^O s_j^O}{2w}.
\end{align}
and obviously, $\sum_{i=1}^{n}\sum_{j=1}^{n} {e_{ij}}=2w$. The table of associated \textit{expected} weights assuming independence between the vertices of a network,  $E={\lbrack e_{ij}\rbrack}_{n\times n}$ can be written as:

\begin{equation}\label{chp5_Tab:adj_E}
\centering
\renewcommand{\arraystretch}{1.2}
\begin{tabular}{ cc| c c c c|c }
&\multicolumn{6}{ c }{} \\
 & & $v_1$ & $v_2$ & $\cdots$ & $v_n$ & $s^E_i$ \\ \cline{2-7}
\multirow{4}{*}{$E={\lbrack e_{ij}\rbrack}_{n\times n}=$}
 & $v_1$ & $e_{11}$ & $e_{12}$ & $\cdots$ & $e_{1n}$ & $s^E_1$ \\
 & $v_2$ & $e_{21}$ & $e_{22}$ & $\cdots$ & $e_{2q}$ & $s^E_2$ \\  	
 & $\vdots$ & $\vdots$ & $\vdots$ 	& $\ddots$ & $\vdots$ & $\vdots$ \\ 
 & $v_n$ & $e_{n1}$ & $e_{n2}$ & $\cdots$ & $e_{nn}$ & $s^E_n$ \\ \cline{2-7}
 & $s_j$ & $s^E_1$ & $s^E_2$ & $\cdots$ & $s^E_n$ & $2w$ \\
\end{tabular}
\end{equation}
\end{definition}

\begin{example}\label{chp5_Ex1:Q-gain_1}
Consider a network with $n=4$ vertices and four edges with a total strength of $w=7$ (Figure~\ref{chp5_fig1:Q-gain_1}). Tables~\ref{chp5_tab_O:Q-gain_1} and \ref{chp5_tab_E:Q-gain_1} provide the tabular representations of the observed, $O$, and expected, $E$, weights  of this network, respectively.
\definecolor{Gray}{gray}{0.85}
\begin{figure}[H]
       \centering
    	\begin{subfigure}[b]{0.25\linewidth}
				\begin{tikzpicture}[inner sep=0pt, minimum size=5mm, auto]
   					\node[node_style,font=\fontsize{9}{10}\selectfont] (v1) at (0,1) {$v_1$};
    				\node[node_style,font=\fontsize{9}{10}\selectfont] (v2) at (1,2) {$v_2$};
	    			\node[node_style,font=\fontsize{9}{10}\selectfont] (v3) at (1,0) {$v_3$};
    				\node[node_style,font=\fontsize{9}{10}\selectfont] (v4) at (2.35,0) {$v_4$};
	    			\draw[edge_style]  (v1) edge node[above,pos=0.3,font=\fontsize{7}{10}\selectfont] {$2$} (v2);
    				\draw[edge_style]  (v1) edge node[below,pos=0.3,font=\fontsize{7}{10}\selectfont] {$1$} (v3);
    				\draw[edge_style]  (v2) edge node[right,pos=0.5,font=\fontsize{7}{10}\selectfont] {$1$} (v3);
    				\draw[edge_style]  (v3) edge node[below,pos=0.5,font=\fontsize{7}{10}\selectfont] {$3$} (v4);
				\end{tikzpicture}
        		\caption{} \label{chp5_fig1:Q-gain_1}
        	\end{subfigure}%
    	\begin{subfigure}[b]{0.33\linewidth}
				\fontsize{8}{10}\selectfont
				\setlength{\tabcolsep}{3.5pt}				
				\renewcommand{\arraystretch}{1.2}			
				\begin{tabular}{ cc| c c c c|c }
				&\multicolumn{6}{ c }{\scriptsize \textup{Observed weights: $O=$}} \\
				 & & $v_1$ & $v_2$ & $v_3$ & $v_4$ & $s_i^O$ \\ \hhline{~*{6}{-}}
				\multirow{4}{*}{}
 					& $v_1$ & \cellcolor{Gray}$0$ & $2$ & $1$& $0$ & $3$ \\
					& $v_2$ & $2$ & \cellcolor{Gray}$0$ & $1$ & $0$ & $3$ \\  	
 					& $v_3$ & $1$ & $1$ & \cellcolor{Gray}$0$ & $3$ & $5$ \\ 
 					& $v_4$ & $0$ & $0$ & $3$ & \cellcolor{Gray}$0$ & $3$ \\ \cline{2-7}
 					& $s_j^O$ & $3$ & $3$ & $5$ & $3$ & $14$ \\
				\end{tabular}
				\normalsize
				\caption{} \label{chp5_tab_O:Q-gain_1}
		\end{subfigure}%
		\begin{subfigure}[b]{0.33\linewidth}
				\fontsize{8}{10}\selectfont
				\setlength{\tabcolsep}{5pt}	
				\renewcommand{\arraystretch}{1.2}			
				\begin{tabular}{ cc| c c c c|c }
				&\multicolumn{6}{ c }{\scriptsize \textup{Expected weights: $E=$}} \\
				 & & $v_1$ & $v_2$ & $v_3$ & $v_4$ & $s_i^E$ \\ \hhline{~*{6}{-}}
				\multirow{4}{*}{}
 					& $v_1$ & \cellcolor{Gray}$\frac{9}{14}$ & $\frac{9}{14}$ & $\frac{15}{14}$ & $\frac{9}{14}$ & $3$ \\
					& $v_2$ & $\frac{9}{14}$ & \cellcolor{Gray}$\frac{9}{14}$ & $\frac{15}{14}$ & $\frac{9}{14}$ & $3$ \\  	
 					& $v_3$ & $\frac{15}{14}$ & $\frac{15}{14}$ & \cellcolor{Gray}$\frac{25}{14}$ & $\frac{15}{14}$ & $5$\\ 
 					& $v_4$ & $\frac{9}{14}$ & $\frac{9}{14}$ & $\frac{15}{14}$ & \cellcolor{Gray}$\frac{9}{14}$ & $3$\\ \cline{2-7}
 					& $s_j^E$ & $3$ & $3$ & $5$ & $3$ & $14$ \\
				\end{tabular}
				\normalsize
				\caption{} \label{chp5_tab_E:Q-gain_1}
		\end{subfigure}	%
\caption[Example of a weighted network and its tabular representations of the \textit{observed} and \textit{expected} weights]{(a) Initial weighted network - $G^W$, (b) tabular representation of the \textit{observed weights} -- $O$, and (c) tabular representation of the \textit{expected weights} -- $E$.}
\end{figure}
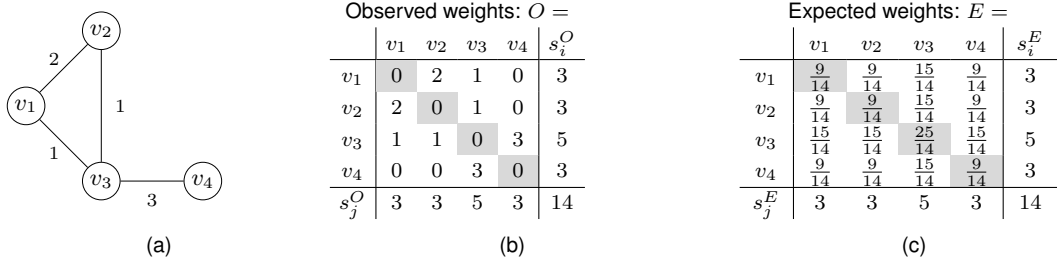
\end{example}

If the normalization factor $\frac{1}{2w}$ is ignored in the definition of weighted modularity~(\ref{weighted_modulatity_Q}) and taking into consideration the above definitions, we may define the modularity of a partition according to the difference between the \textit{observed} and the \textit{expected} edge weights as follows.

\begin{definition}[Modularity for the difference between the observed and the expected weights]
Given an undirected weighted network $G^W$ and a partition $\mathcal{C}=\{C_1,C_2,...,C_q\}$ of its vertices into $q$ sets, the (new) unstandardised weighted modularity $Q^N$ of partition $\mathcal{C}$ is defined as:
\begin{equation}\label{chp5_Eq:Q_vertices}
Q^N=\sum_{i=1}^n \sum_{j=1}^n \left(o_{ij}-e_{ij}\right) \delta(C_i, C_j)=\sum_{i=1}^n \sum_{j=1}^n \left(o_{ij}-\frac{s^O_i s^O_j}{2w}\right) \delta(C_i, C_j),
\end{equation}
\noindent
which may also be written as a \textup{sum over all the different communities in the community structure}, $\mathcal{C}$:
\begin{equation}\label{chp5_Eq:Q_communities}
Q^N=\sum_{C\in \mathcal{C}} \sum_{i,j\in C} \left(o_{ij}-e_{ij}\right).
\end{equation}
\end{definition}

Likewise, Newmans' normalization of the modularity for unweighted networks, also known as an \textit{assortativity coefficient} (for details, see~\citealp{Newman:2010wp}), may be extended to the case of weighted networks. The \textit{normalized modularity} for weighted networks $Q_{norm}$ is given by:
\begin{equation}\label{chp4:Eq_Qnorm-2}
Q^N_{norm}=\frac{Q^N}{Q^N_{max}}=\frac{\sum_{C\in \mathcal{C}} \sum_{i,j\in C} \left(o_{ij}-e_{ij}\right)}{2w-\sum_{C\in \mathcal{C}} \sum_{i,j\in C} e_{ij}}
\end{equation}

\subsubsection{Maximization of modularity -- The Louvain algorithm}
\label{louvain_modularity_maximization}

The methodology we use to maximize the modularity is the so-called Louvain algorithm~\citep{Blondel:2008do}. This algorithm is characterized by the following: initially each vertex forms a community, then for each \textit{pass} of the algorithm there are two \textit{phases}, at the 1\textsuperscript{st} phase (\textit{optimization}), modularity $(Q^N)$ and modularity gain $(\Delta Q^N=Q^N_{new} - Q^N_{last})$ are iteratively computed for all vertices in a local greedy approach until no movement of a vertex from  its original community yields a gain in modularity; at the 2\textsuperscript{nd} phase, the \textit{aggregation of the network} is done by summing the weights for the formed communities. Each \textit{pass} of the algorithm is repeated until convergence, i.e., until modularity cannot be increased. This iterative procedure produces one partition per pass, thus creating a hierarchy of communities.

Therefore, to evaluate the change in modularity for weighted networks obtained by merging two communities $C_r$ and $C_s$ into a single community $C_t=C_r \cup C_s$, we may compare the modularity before and after the merge:

\begin{equation} \label{chp5_Eq:DeltaQ_1}
\Delta Q^N=Q^N_{new} - Q^N_{last}
\end{equation}
where $Q^N_{new}$ and $Q^N_{last}$ are respectively the modularity after and before the merging of $C_r$ with $C_s$.\\

However, since merging a pair of communities between which there are no edges cannot increase the modularity, we need only compute the change in modularity for pairs of connected communities as~\citep{Newman:2004ws}:
\begin{align}
\Delta Q^N &= \left[o_{rr}+o_{ss}+o_{rs}+o_{sr}-\left(e_{rr}+e_{ss}+e_{rs}+e_{sr}\right)\right]-\left[o_{rr}-e_{rr}\right]-\left[o_{ss}-e_{ss}\right] \nonumber\\
			&= \cancel{o_{rr}}+\cancel{o_{ss}}+o_{rs}+o_{sr}-\cancel{e_{rr}}-\cancel{e_{ss}}-e_{rs}-e_{sr}-\cancel{o_{rr}}+\cancel{e_{rr}}-\cancel{o_{ss}}+\cancel{e_{ss}} \nonumber\\
\Delta Q^N &=o_{rs}+o_{sr}-e_{rs}-e_{sr} \nonumber\\
\Delta Q^N &=2\left(o_{rs}-e_{rs}\right),  \label{chp5_Eq:DeltaQ_2}
\end{align}
where $o_{rs}$ and and $e_{rs}$ are respectively the observed and expected weights of edges connecting vertices in community $r$ to vertices in community $s$. This reduced formulation of modularity gain~(\ref{chp5_Eq:DeltaQ_2}) is computationally more efficient than its initial expression~(\ref{chp5_Eq:DeltaQ_1}), because it acts locally and not globally~\citep{Clauset:2004uy}.

The following Table~\ref{chp5_Sec_1.1_tab:Gain} illustrates the calculation of the modularity gain and the process of placing a vertex in another neighbouring community, for a very small weighted network (triplet). In order to distinguish the two formulations of the modularity gain presented above, we will denote the former~(\ref{chp5_Eq:DeltaQ_1}) as $\Delta Q^N_1$ and the latter, its reduced formulation~(\ref{chp5_Eq:DeltaQ_2}), as $\Delta Q_2^N$.

\begin{table}[H]
\caption{Modularity gain by moving vertex $v_1$ to neighbouring communities of vertices $v_2$ and $v_3$.}
\label{chp5_tab:Q-gain}
\label{chp5_Sec_1.1_tab:Gain}
\footnotesize
\centering
\begin{adjustbox}{max width=\textwidth}
\renewcommand{\arraystretch}{1.4}
\begin{tabular}{p{3.5cm}|l|l}
\hline
& \multicolumn{2}{c}{Placing vertex $v_1$ in neighbouring communities $v_2$ and $v_3$} \\[0.3ex] \cline{2-3}
&
\multicolumn{1}{l|}{\begin{tikzpicture}[inner sep=0pt, minimum size=5mm, auto, remember picture]
   	\node[node_style,,font=\fontsize{9}{10}\selectfont,fill=gray!20] (v1) at (0,1) {$v_1$};
    \node[node_style,,font=\fontsize{9}{10}\selectfont,fill=gray!20] (v2) at (1,2) {$v_2$};
	\node[node_style,font=\fontsize{9}{10}\selectfont] (v3) at (1,0) {$v_3$};
	\node [rotate=-135][draw,dashed,inner sep=0pt, circle,yscale=.5, fit={(v1) (v2)}] {};
	\node[] (v2') at (-0.15,2) {$v_{2'}$};  
    \node[fit=(current bounding box),inner ysep=2mm,inner xsep=2mm]{};  
	\draw[edge_style]  (v1) edge node[above,pos=0.3,font=\fontsize{7}{10}\selectfont] {$2$} (v2);
   	\draw[edge_style]  (v1) edge node[below,pos=0.3,font=\fontsize{7}{10}\selectfont] {$1$} (v3);
\hspace{0.1cm}
	\node[node_style,fill=gray!20,minimum size=.8cm,font=\fontsize{8}{7}\selectfont,inner sep=0] (v2x) at (3,1) {$v_{2'}$};
	\node[node_style,font=\fontsize{9}{10}\selectfont] (v3x) at (4,0) {$v_3$};
	\node[] (v*) at (1.5,1) {};  
	\draw[edge_style,loop above]  (v2x) edge node[above,pos=0.5,font=\fontsize{8}{10}\selectfont] {$4$} (v2x);
	\draw[edge_style]  (v2x) edge node[below,pos=0.3,font=\fontsize{7}{10}\selectfont] {$1$} (v3x);
	\draw[thick, -latex, shorten <= -15pt, shorten >= 15pt] (v*) -- node [] {} (v2x);
\end{tikzpicture}}
&
\multicolumn{1}{l}{\begin{tikzpicture}[inner sep=0pt, minimum size=5mm, auto, remember picture]
   	\node[node_style,,font=\fontsize{9}{10}\selectfont,fill=gray!20] (v1) at (0,1) {$v_1$};
    \node[node_style,font=\fontsize{9}{10}\selectfont] (v2) at (1,2) {$v_2$};
	\node[node_style,,font=\fontsize{9}{10}\selectfont,fill=gray!20] (v3) at (1,0) {$v_3$};
	\node [rotate=135][draw,dashed,inner sep=0pt, circle,yscale=.5, fit={(v1) (v3)}] {};
	\node[] (v3') at (1.15,1) {$v_{3'}$};  
	\draw[edge_style]  (v1) edge node[above,pos=0.3,font=\fontsize{7}{10}\selectfont] {$2$} (v2);
   	\draw[edge_style]  (v1) edge node[below,pos=0.3,font=\fontsize{7}{10}\selectfont] {$1$} (v3);
\hspace{0.4cm}
	\node[node_style,fill=gray!20,minimum size=.8cm,font=\fontsize{8}{7}\selectfont,inner sep=0] (v3xx) at (3,1) {$v_{3'}$};
	\node[node_style,font=\fontsize{9}{10}\selectfont] (v2xx) at (4,2) {$v_2$};
	\node[] (v**) at (1.5,1) {};  
	\draw[edge_style,loop below]  (v3xx) edge node[below,pos=0.5,font=\fontsize{8}{10}\selectfont] {$2$} (v3xx);
	\draw[edge_style]  (v2xx) edge node[above,pos=0.8,font=\fontsize{7}{10}\selectfont] {$2$} (v3xx);
	\draw[thick, -latex, shorten <= -15pt, shorten >= 15pt] (v**) -- node [] {} (v3xx);
\end{tikzpicture}}
\\
Modularity Gain -- $\Delta Q^N$ & \multicolumn{1}{c|}{$\Delta Q^N_{v_1\to v_2}$} & \multicolumn{1}{c}{$\Delta Q^N_{v_1\to v_3}$}
\\[0.5ex] \hline
\rowcolor{gray!20}$1.\ \Delta Q_1^N=Q^N_{new} - Q^N_{last}$ & $\Delta Q^N_{v_1\to v_2}=-\frac{2}{6}+\frac{14}{6}=2$ & $\Delta Q^N_{v_1\to v_3}=-\frac{8}{6}+\frac{14}{6}=1$ \\ [0.5ex] \hline
\hspace{1.3em}\color{black!90}$Q^N_{last}$ & \color{black!90}$Q^N_{last}=(0-\frac{9}{6})+(0-\frac{4}{6})+(0-\frac{1}{6})=-\frac{14}{6}$ & \color{black!90}$Q^N_{last}=(0-\frac{9}{6})+(0-\frac{4}{6})+(0-\frac{1}{6})=-\frac{14}{6}$ \\
\hspace{1.3em}\color{black!90}$Q^N_{new}$ & \color{black!90}$Q^N_{new}=(4-\frac{25}{6})+(0-\frac{1}{6})=-\frac{2}{6}$ & \color{black!90}$Q^N_{new}=(2-\frac{16}{6})+(0-\frac{4}{6})=-\frac{8}{6}$ \\[0.5ex] \hline
\rowcolor{gray!20}$2.\ \Delta Q_2^N=2\left(o_{rs}-e_{rs}\right)$ & $\Delta Q^N_{v_1\to v_2}=2\left(o_{12}-e_{12}\right)=2\times \left(2-\frac{6}{6}\right)=2$ & $\Delta Q^N_{v_1\to v_3}=2\left(o_{13}-e_{13}\right)=2\times \left(1-\frac{3}{6}\right)=1$ \\ [0.5ex] \hline
\scriptsize{\textbf{Decision} $\Rightarrow$ choose the \textit{maximum} gain in modularity} & \multicolumn{2}{c}{\multirow{2}{*}{$\Delta Q^N_{v_1\to v_2}>\Delta Q^N_{v_1\to v_3}$ $\Rightarrow$ Place vertex $v_1$ with $v_2$}} \\
\hline
\end{tabular}
\end{adjustbox}
\end{table}


\section{Community Detection in Interval-Weighted Networks based on a Contingency Table}
\label{CD_IWN}

Based on the definitions discussed above in Section~\ref{CD_WN}, we now extend modularity and modularity gain to the case of IWN. However, due to the above mentioned problems intrinsic to interval arithmetic (e.g. interval dependency, among others -- see Section~\ref{Subsection_2.1}) the straightforward extensions were not achieved. We also note that when using an IWN, an adjustment is required when performing expected frequency calculations. These difficulties lead us to designing a whole new approach to solve the problem. First, we define several measures either to evaluate the difference between two intervals (Subsection~\ref{chp5_SubSec_IntervalDifferences}), or to calculate the modularity and respective modularity gain for interval-weighted networks (Subsection~\ref{chp5_SubSec_IntervalModularity}). Then we develop two different strategies adapted to deal with IWN, to optimize the modularity according to the Louvain algorithm, called: ``Method 1: Classic Louvain (CL)'' (Section~\ref{chp5_subsction_CL--Method 1: Intervals Sum}) and ``Method 2: Hybrid Louvain (HL)'' (Section~\ref{chp5_subsction_CL--Method 2: Intervals Min--Max}). These different approaches allow obtaining solutions according to different criteria, such as capturing different variability from data.

\subsection{Modularity for interval-weighted networks}
\label{chp5:sec_2}

In Section~\ref{louvain_modularity_maximization}, we have shown that the evaluation of the modularity gain $\Delta Q^N$ in weighted networks when moving an isolated vertex $i$ into a community $C_j$ (Louvain's optimization phase) is a twofold process: (i) calculating the difference between the modularity of the network \textit{after} and \textit{before} the vertex $i$ is removed from its community and is placed in the neighbouring community $j$, using~(\ref{chp5_Eq:DeltaQ_1}) or its reduced formulation~(\ref{chp5_Eq:DeltaQ_2}) and; (ii) inserting vertex $i$ into  community $C_j$ only if that change increases the value of network modularity $(\Delta Q^N>0)$.
However, in directly extending the calculation formulas of modularity and modularity gain to interval data, we face two major setbacks: (i) the first is related to the interval arithmetic pitfalls, such as, the \textit{interval dependency} (e.g. $[-2,2]-[-2,2]=[-4,4]$ and not $[0,0]$) or because only a \textit{subdistributive law} is valid $\left(\left(\exists\ X,Y,Z \in [\mathbb{R}]\right)\ \left(Z \times (X+Y) \subseteq Z \times X + Z \times Y\right)\right)$ (see Section~\ref{Subsection_2.1} for details); (ii) the second is because one way of evaluating the difference between two intervals is to use a measure of distance, however \textit{a distance is always non-negative}. There are several distance measures in the literature to compare two intervals, one of them is the \textit{Hausdorff distance}~\citep{Bryant:1985wi,billard2006symbolic}: $d_H\big(\interval{\underline{x}}{\overline{x}}, \interval{\underline{y}}{\overline{y}}\big)=\max\big(|\underline{x}-\underline{y}|,|\overline{x}-\overline{y}|\big)$. Still, by using a distance, it means by definition that the outcome value of both modularity and modularity gain are always non-negative, which makes it impossible to determine if a vertex stays in its own community or moves to a neighbourhood community (see Section~\ref{louvain_modularity_maximization}).

\subsubsection{Interval difference -- $D$}
\label{chp5_SubSec_IntervalDifferences}
Due to these characteristics of interval arithmetic that prevent us from realizing a direct extension of the previous modularity and modularity gain measures, to evaluate the difference between two intervals, we propose a measure $D$, which is based on the \textit{Hausdorff distance} but it \textit{does take into account the \textbf{sign}} of the highest value, to evaluate the difference between two intervals:

\begin{definition}[Difference -- $D$]\label{chp5:Def_d2}
The difference $D$ between two intervals $\interval{\underline{x}}{\overline{x}}$ and $\interval{\underline{y}}{\overline{y}}$ is defined to be
\begin{equation}\label{chp5:Eq_d2}
D\big([\underline{x},\overline{x}],[\underline{y},\overline{y}]\big)=\max\big\{|\underline{x}-\underline{y}|,|\overline{x}-\overline{y}|\big\}\times sign\ argmax\big\{|\underline{x}-\underline{y}|,|\overline{x}-\overline{y}|\big\}
\end{equation}
\end{definition}

\begin{example}\label{chp5:Ex_metrics1}
Let $X=[\underline{x},\overline{x}]$ and $Y=[\underline{y},\overline{y}]$ be a pair of arbitrary intervals ($X,Y \subseteq \mathbb{R^+}$). Below we show different  calculations of the difference $D$ for three types of intervals:
\begin{itemize}
\setlength\itemsep{2pt}
\item Non-overlapping intervals: $X=[1,3]$ and $Y=[4,5]$\\
$D(X,Y)=\max\{|1-4|,|3-5|\}\times (-1)=\max\{3,2\}\times (-1)=-3$
\item Partially overlapping intervals: $X=[2,5]$ and $Y=[1,3]$\\
$D(X,Y)=\max\{|2-1|,|5-3|\}\times (+1)=\max\{1,2\}\times (+1)=+2$
\item Completely overlapping intervals: $X=[1,5]$ and $Y=[3,4]$\\
$D(X,Y)=\max\{|1-3|,|5-4|\}\times (-1)=\max\{2,1\}\times (-1)=-2$
\end{itemize}
\end{example}

Based on measure $D$, we develop a working framework, which we call ~\textit{Interval Modularity $(Q^I)$} that extends the ``classical'' modularity and the modularity gains of weighted networks to the case of interval-weighted networks.

\begin{note}
Before moving on to the following generalizations, it is important to note that in interval arithmetic, the difference $D$ between the sum of intervals (see Section~\ref{Subsection_2.1}) is different from the sum of the differences between those intervals (e.g., considering four intervals $X$, $Y$, $X'$, and $Y'$: $D(X+Y, X'+Y')\neq D(X,X')+D(Y,Y'))$~\citep{Moore:2009uc}. Thus, the derivation of the modularity gain $\Delta Q_1^N=Q^N_{new}-Q^N_{last}$~(\ref{chp5_Eq:DeltaQ_1}) into $\Delta Q_2^N=2\left(o_{rs}-e_{rs}\right)$~(\ref{chp5_Eq:DeltaQ_2}), with the use of intervals instead of real numbers \textit{is not verified}, i.e., in general, $\Delta Q^I_{1}\neq \Delta Q^I_{2}$. 
\end{note}

\subsubsection{Interval Modularity -- $Q^I$}
\label{chp5_SubSec_IntervalModularity}

Given an undirected interval-weighted network $G^I$ and a partition $\mathcal{C}=\{C_1,\allowbreak C_2,\allowbreak\dots,\allowbreak C_q\}$ of its vertices into $q$ sets, the generalization of modularity $Q^N$~(\ref{chp5_Eq:Q_communities}), modularity gain $\Delta Q^N$~(\ref{chp5_Eq:DeltaQ_1}) and the normalized modularity $Q_{norm}$~(\ref{chp4:Eq_Qnorm-2}) to \textit{interval data} is done as follows:
\begin{definition} [Modularity for interval-weighted networks -- $Q^I\ (Q^I\in \mathbb{R})$]
\begin{equation}
Q^I=\sum_{r=1}^q D\left(o_{rr},e_{rr}\right)
\end{equation}
where ``D'' represents the difference between the observed $o_{rr}$ and the expected $e_{rr}$ interval-weights of community $r$ (see (\ref{chp5:Eq_d2})).
\end{definition}

Likewise, assuming that we have a fixed partition consisting in two communities $C_r$ and $C_s$, to evaluate the modularity gain resulting from the merging of $C_r$ and $C_s$ into a single community $C_t=C_r \cup C_s$, the modularity gain for interval-weighted networks is defined as follows:

\begin{definition} [Modularity gain for interval-weighted networks -- $\Delta Q^I\ (\Delta Q^I\in \mathbb{R}$]
\begin{equation}
\Delta Q^I=Q^I_{new} - Q^I_{last}
\end{equation}
\end{definition}

In the same way that we made the straightforward extension of both the modularity and modularity gain of weighted networks to interval-weighted networks, we will proceed to the \textit{normalization of modularity} in the case of interval-weighted networks. Using~(\ref{chp4:Eq_Qnorm-2}), we obtain

\begin{definition} [Normalized modularity for interval-weighted networks -- $Q^I_{norm}\ (Q^I_{norm}\in \mathbb{R}$]
Considering the reduced formula of interval-weighted modularity,
\begin{equation}
Q^I_{norm}=\frac{Q^I}{Q^I_{\max}}=\frac{\sum_r D(o_{rr},e_{rr})}{D\left([2\underline{w},2\overline{w}],\sum_r e_{rr}\right)}\\
\end{equation}
\end{definition}

\subsection{Methodology}
\label{chp5_SubSec:Methodology}

\begin{figure}[h!]
	\centering
    \includegraphics[scale=0.5, clip, trim={0cm 11cm 0cm 11cm}]{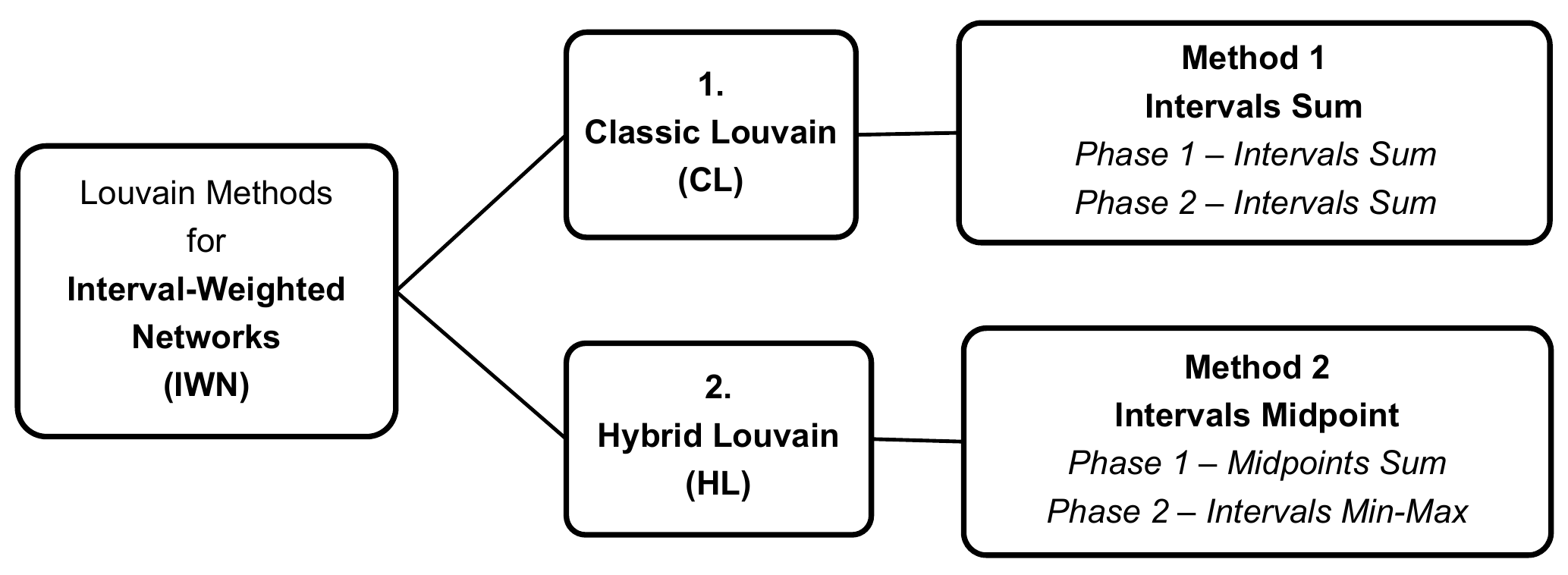}
     \caption[Sketch of the Louvain method extended to interval-weighted networks developed in this thesis]{Sketch of the Louvain method extended to interval-weighted networks, ``Method 1'' and ``Method 2''.}
     \label{chp5_fig:Esquema_Louvain}
\end{figure}

We aim at applying the well-known Louvain algorithm for community detection to networks whose values (or weights) of the connections between the vertices are represented by intervals instead of real values (``interval-weighted networks''). The implementation of this methodology to interval-weighted networks is accomplished through the design of two new approaches that we name \textit{Classic Louvain (CL)} and \textit{Hybrid Louvain (HL)}, which in turn may consider two different methods: (i) ``Method 1: Intervals Sum'' and; (ii) ``Method 2: Intervals Midpoint''.\\
Figure~\ref{chp5_fig:Esquema_Louvain} depicts an illustrative scheme for each of these approaches and their respective methods.

Next, we describe each of the methods, proposing the extension to interval data of the definitions developed and presented in Section~\ref{chp5:sec_1}. We use the representation of an \textit{interval-weighted network} in the form of an \textit{interval-weighted matrix}, and then propose a new approach to extend the modularity, modularity gain and consequently the Louvain algorithm to community detection in interval-weighted networks. As mentioned above, the two different approaches for the ``classic'' Louvain algorithm are:

\begin{itemize}[leftmargin=1cm,label={{\large \textbullet}}]
\setlength\itemsep{0.15cm}
\item the first one, baptised as \textbf{``Method 1: Intervals Sum''} follows \citet{Blondel:2008do} procedure, i.e., both the \textit{optimization} on the 1\textsuperscript{st} phase and the \textit{aggregation of the network} on the 2\textsuperscript{nd} phase, are accomplished by \textit{summing} the intervals (henceforward, ``Method 1'');
\item in the second one, named \textbf{``Method 2: Intervals Midpoint''}, the \textit{optimization} on the 1\textsuperscript{st} phase is performed by \textit{using the midpoints} of the intervals, while on the 2\textsuperscript{nd} phase, the \textit{aggregation of the network} is done by selecting the minimum and the maximum values of the intervals for the formed communities, in order to capture the maximum variability present (henceforward, ``Method 2'').
\end{itemize}

\subsection{Classic Louvain -- Method 1: Intervals Sum}
\label{chp5_subsction_CL--Method 1: Intervals Sum}

Using intervals to represent weighted network data, we obtain an interval-weighted table~\citep{Hu:2008if,Moore:2009uc}. In order to follow the notation adopted so far in this manuscript, we consider that the intervals formed by the lower and upper values between any two vertices, $[\underline{w}_{ij},\overline{w}_{ij}]$ will instead be denoted as the lower and upper values of the observed weights between vertices $[\underline{o}_{ij},\overline{o}_{ij}]$.

\begin{definition} [Contingency table for the observed interval-weights -- $O^I$] \label{chp5_Def:adj_O2}
A contingency table whose cells represent the \textup{observed interval-weights} $o^I_{ij}=[\underline{o}_{ij},\overline{o}_{ij}]$ $(\overline{o}_{ij} \geqslant \underline{o}_{ij}>0;\ o^I_{ij} \subseteq \mathbb{R^+})$, if there is an weighted edge between vertices $(i,j)$, and zero otherwise, is called an interval contingency table, denoted by $O^I$. The \textup{interval marginal sums} for each row or column and the \textup{interval total weight} or \textup{interval strength} attached (or linked) to vertex $i$, are denoted by $s^{IO}_i=\sum_{j=1}^n [\underline{o}_{ij},\overline{o}_{ij}]$, and the total weight is $\sum_{i=1}^{n} s^{IO}_i = \sum_{j=1}^{n} s^{IO}_j = \sum_{i=1}^{n}\sum_{j=1}^{n} [\underline{o}_{ij},\overline{o}_{ij}]$. Thus, the table of associated \textit{observed} interval weights can be represented as:

\begin{equation}\label{chp5_Tab:adj_O2}
\centering
\setlength{\tabcolsep}{1.5pt}
\setlength{\extrarowheight}{2.2pt}		
\begin{tabular}{ cc| c c c c|c }
&\multicolumn{6}{ c }{} \\
 & & $v_1$ & $v_2$ & $\cdots$ & $v_n$ & $s^{IO}_i$ \\ \cline{2-7}
\multirow{4}{*}{$O^I{=}o^I_{ij}{=}\Big[\underline{o}_{ij},\overline{o}_{ij}\Big]_{n\times n}{=}$}
 & $v_1$ & $[\underline{o}_{11},\overline{o}_{11}]$ & $[\underline{o}_{12},\overline{o}_{12}]$ & $\cdots$ & $[\underline{o}_{1n},\overline{o}_{1n}]$ & $\sum\limits_{j=1}^n [\underline{o}_{1j},\overline{o}_{1j}]$ \\
 & $v_2$ & $[\underline{o}_{21},\overline{o}_{21}]$ & $[\underline{o}_{22},\overline{o}_{22}]$ & $\cdots$ & $[\underline{o}_{2n},\overline{o}_{2n}]$ & $\sum\limits_{j=1}^n [\underline{o}_{2j},\overline{o}_{2j}]$ \\  	
 & $\vdots$ & $\vdots$ & $\vdots$ 	& $\ddots$ & $\vdots$ & $\vdots$ \\ 
 & $v_n$ & $[\underline{o}_{n1},\overline{o}_{n1}]$ & $[\underline{o}_{n2},\overline{o}_{n2}]$ & $\cdots$ & $[\underline{o}_{nn},\overline{o}_{nn}]$ & $\sum\limits_{j=1}^n [\underline{o}_{nj},\overline{o}_{nj}]$ \\ \cline{2-7}
 & $s^{IO}_j$ & $\sum\limits_{i=1}^n [\underline{o}_{i1},\overline{o}_{i1}]$ & $\sum\limits_{i=1}^n [\underline{o}_{i2},\overline{o}_{i2}]$ & $\cdots$ & $\sum\limits_{i=1}^n [\underline{o}_{in},\overline{o}_{in}]$ & $\sum\limits_{i=1}^{n}\sum\limits_{j=1}^{n} [\underline{o}_{ij},\overline{o}_{ij}]$
\end{tabular}
\end{equation}

In order to simplify future notations, the \textup{interval marginal sums} will be denoted as $s^{IO}_i=\left[\underline{s}^{IO}_i,\overline{s}^{IO}_i\right]$. Likewise, the total interval-weight will be denoted as $[2\underline{w},2\overline{w}]$.
\end{definition}

Analogously to Definition~\ref{chp5_Def:adj_E}, the \textup{expected interval-weights} of the interval contingency table, are defined as follows.

\begin{definition}
Denoting the \textup{expected interval-weights} \textup{assuming independence} as $e^I_{ij}$, the interval-weight that would be obtained if the hypothesis of row-column independence were true, and each element calculated by expression~(\ref{chp5_Eq:interval_expected-freq}) following specific mathematical operations for interval division (as defined previously in Subsection~\ref{Subsection_2.1}), we obtain

\begin{align}\label{chp5_Eq:interval_expected-freq}
e^I_{ij}&=s^{IO}_i{\times} s^{IO}_j{\times} \Bigg[\frac{1}{2\overline{w}},\frac{1}{2\underline{w}}\Bigg]{=}\Big[\underline{s}^{IO}_i,\overline{s}^{IO}_i\Big]{\times} \Big[\underline{s}^{IO}_j,\overline{s}^{IO}_j\Big]{\times} \Bigg[\frac{1}{2\overline{w}},\frac{1}{2\underline{w}}\Bigg]{=}\Bigg[\frac{\underline{s}^{IO}_i \underline{s}^{IO}_j}{2\overline{w}},\frac{\overline{s}^{IO}_i \overline{s}^{IO}_j}{2\underline{w}}\Bigg]\\
&(0\notin [2\underline{w},2\overline{w}])\nonumber
\end{align}
\end{definition}

The contingency table for the \textit{expected interval-weights} assuming independence between the vertices $E^I={\lbrack e_{ij}\rbrack}_{n\times n}$, is represented as:

\begin{equation}\label{chp5_Tab:adj_O2}
\centering
\setlength{\extrarowheight}{3pt}		
\begin{tabular}{ cc| c c c c|c }
 & & $v_1$ & $v_2$ & $\cdots$ & $v_n$ &  \\ \cline{2-7}
\multirow{4}{*}{$E^I={e^I_{ij}=[\underline{e}_{ij},\overline{e}_{ij}]}_{n\times n}=$}
 & $v_1$ & $[\underline{e}_{11},\overline{e}_{11}]$ & $[\underline{e}_{12},\overline{e}_{12}]$ & $\cdots$ & $[\underline{e}_{1n},\overline{e}_{1n}]$ &  \\
 & $v_2$ & $[\underline{e}_{21},\overline{e}_{21}]$ & $[\underline{e}_{22},\overline{e}_{22}]$ & $\cdots$ & $[\underline{e}_{2n},\overline{e}_{2n}]$ &  \\  	
 & $\vdots$ & $\vdots$ & $\vdots$ 	& $\ddots$ & $\vdots$ &  \\ 
 & $v_n$ & $[\underline{e}_{n1},\overline{e}_{n1}]$ & $[\underline{e}_{n2},\overline{e}_{n2}]$ & $\cdots$ & $[\underline{e}_{nn},\overline{e}_{nn}]$ & \\ \cline{2-7}
&  &  &  &  &  &  
\end{tabular}
\end{equation}

\begin{note}\label{chp5_Note:Expected_marginals}
The Contingency tables for the expected interval-weights $(E^I)$ do not have the row and column marginal totals, as well as the table total (see Example~\ref{chp5_Ex:interval_tables} below), since due to the mathematical operations for interval division, these totals no longer correspond to the totals of the contingency table of observed values. As these totals are not used in our mathematical procedures, for the sake of simplicity, we have chosen not to show them in the table.
\end{note}

\begin{example}\label{chp5_Ex:interval_tables}
Consider an interval-weighted network $G^I$ with four vertices $n=4$ and four edges with a total strength of $w=[5,9]$ (Figure~\ref{chp5_fig:interval-net}). Tables (b) (Figure~\ref{chp5_tab_O:Q-gain_1}) and (c) (Figure~\ref{chp5_tab_E:Q-gain_1}) correspond to the tabular representations of the observed, $O^I$, and expected, $E^I$, interval-weights  of this network, respectively. To serve as an example of what was described in Note~\ref{chp5_Note:Expected_marginals}, exceptionally, the row and column marginal totals, as well as the table total, are shown.
\definecolor{Gray}{gray}{0.85}
\begin{figure}[H]
       \centering
    	\begin{subfigure}[c]{0.99\linewidth}
				\centering        
				\begin{tikzpicture}[inner sep=0pt, minimum size=5mm, auto]
   					\node[node_style,font=\fontsize{9}{10}\selectfont] (v1) at (0,1) {$v_1$};
    				\node[node_style,font=\fontsize{9}{10}\selectfont] (v2) at (1,2) {$v_2$};
	    			\node[node_style,font=\fontsize{9}{10}\selectfont] (v3) at (1,0) {$v_3$};
    				\node[node_style,font=\fontsize{9}{10}\selectfont] (v4) at (2.35,0) {$v_4$};
	    			\draw[edge_style]  (v1) edge node[above,sloped,pos=0.5,font=\fontsize{7}{10}\selectfont] {$[1,3]$} (v2);
    				\draw[edge_style]  (v1) edge node[below,sloped,pos=0.5,font=\fontsize{7}{10}\selectfont] {$[1,1]$} (v3);
    				\draw[edge_style]  (v2) edge node[right=0.1,pos=0.5,font=\fontsize{7}{10}\selectfont] {$[1,1]$} (v3);
    				\draw[edge_style]  (v3) edge node[below,pos=0.5,font=\fontsize{7}{10}\selectfont] {$[2,4]$} (v4);
				\end{tikzpicture}
        		\caption{} \label{chp5_fig:interval-net}
        	\end{subfigure}%
        	             	
    	\begin{subfigure}[c]{0.45\linewidth}
    		\setlength{\tabcolsep}{3.5pt}
    		\setlength{\extrarowheight}{3.5pt}
				\centering
				\fontsize{8}{10}\selectfont
				\begin{tabular}{ cc| c c c c|c }
				&\multicolumn{6}{ c }{\footnotesize \textup{Observed interval-weights: $O^I=$}} \\
				 & & $v_1$ & $v_2$ & $v_3$ & $v_4$ & $s_i^{IO}$ \\ \hhline{~*{6}{-}}
				\multirow{4}{*}{}
 					& $v_1$ & \cellcolor{Gray}$[0,0]$ & $[1,3]$ & $[1,1]$ & $[0,0]$ & $[2,4]$ \\
					& $v_2$ & $[1,3]$ & \cellcolor{Gray}$[0,0]$ & $[1,1]$ & $[0,0]$ & $[2,4]$ \\  	
 					& $v_3$ & $[1,1]$ & $[1,1]$ & \cellcolor{Gray}$[0,0]$ & $[2,4]$ & $[4,6]$ \\ 
 					& $v_4$ & $[0,0]$ & $[0,0]$ & $[2,4]$ & \cellcolor{Gray}$[0,0]$ & $[2,4]$ \\ \cline{2-7}
 					& $s_j^{IO}$ & $[2,4]$ & $[2,4]$ & $[4,6]$ & $[2,4]$ & $[10,18]$ \\
				\end{tabular}
				\normalsize
				\caption{} \label{chp5_tab_O:Q-gain_1}
		\end{subfigure}%
		\begin{subfigure}[c]{0.55\linewidth}
			\setlength{\tabcolsep}{2pt}
			\setlength{\extrarowheight}{3pt}		
				\centering
				\fontsize{8}{10}\selectfont
				\begin{tabular}{ cc| c c c c|c }
				&\multicolumn{6}{ c }{\footnotesize \textup{Expected interval-weights: $E^I=$}} \\
				 & & $v_1$ & $v_2$ & $v_3$ & $v_4$ & $s_i^{IE}$ \\ \hhline{~*{6}{-}}
				\multirow{4}{*}{}
 					& $v_1$ & \cellcolor{Gray}$\Big[\frac{4}{18},\frac{16}{10}\Big]$ & $\Big[\frac{4}{18},\frac{16}{10}\Big]$ & $\Big[\frac{8}{18},\frac{24}{10}\Big]$ & $\Big[\frac{4}{18},\frac{16}{10}\Big]$ & $\Big[\frac{20}{18},\frac{72}{10}\Big]$ \\
					& $v_2$ & $\Big[\frac{4}{18},\frac{16}{10}\Big]$ & \cellcolor{Gray}$\Big[\frac{4}{18},\frac{16}{10}\Big]$ & $\Big[\frac{8}{18},\frac{24}{10}\Big]$ & $\Big[\frac{4}{18},\frac{16}{10}\Big]$ & $\Big[\frac{20}{18},\frac{72}{10}\Big]$ \\  	
 					& $v_3$ & $\Big[\frac{8}{18},\frac{24}{10}\Big]$ & $\Big[\frac{8}{18},\frac{24}{10}\Big]$ & \cellcolor{Gray}$\Big[\frac{16}{18},\frac{36}{10}\Big]$ & $\Big[\frac{8}{18},\frac{24}{10}\Big]$ & $\Big[\frac{40}{18},\frac{108}{10}\Big]$\\ 
 					& $v_4$ & $\Big[\frac{4}{18},\frac{16}{10}\Big]$ & $\Big[\frac{4}{18},\frac{16}{10}\Big]$ & $\Big[\frac{8}{18},\frac{24}{10}\Big]$ & \cellcolor{Gray}$\Big[\frac{4}{18},\frac{16}{10}\Big]$ & $\Big[\frac{20}{18},\frac{72}{10}\Big]$\\ \cline{2-7}
 					& $s_j^{IE}$ & $\Big[\frac{20}{18},\frac{72}{10}\Big]$ & $\Big[\frac{20}{18},\frac{72}{10}\Big]$ & $\Big[\frac{40}{18},\frac{108}{10}\Big]$ & $\Big[\frac{20}{18},\frac{72}{10}\Big]$ & $\Big[\frac{100}{18},\frac{324}{10}\Big]$ \\
				\end{tabular}
				\normalsize
				\caption{} \label{chp5_tab_E:Q-gain_1}
		\end{subfigure}	%
\caption[Example of an weighted network and its tabular representations of the \textit{observed} and \textit{expected} weights (Method 1)]{(a) Interval-weighted network - $G^I$, (b) tabular representation of the \textit{observed interval-weights} -- $O^I$, and (c) tabular representation of the \textit{expected interval-weights} -- $E^I$.}
\end{figure}
\end{example}

\subsubsection{Adjustments of the Expected interval-weights [Method 1]}
\label{chp5_Subsec_adjustments_1}

When calculating the expected frequencies according to~(\ref{chp5_Eq:interval_expected-freq}), it should be noted that the value corresponding to the total weight for each of these expected frequencies must pass through an \textit{``adjustment''} of its lower $(2\underline{w})$ and upper limits $(2\overline{w})$. 

This is done because, when calculating the interval corresponding to the expected frequency of each pair of vertices of the network $(e^I_{ij})$, when both limits of the intervals of these vertex pairs $(i,j)$ are at the minimum possible value, the maximum value of the corresponding interval for the network weight is never achieved. Likewise, when both limits of the intervals of the vertex pairs are at the maximum possible value, the minimum value of the interval corresponding to the total weight of the network is never reached. Obviously, these adjustments cause a reduction in the width of the total interval-weight for each pair of vertices of the contingency table. Thus, new \textit{expected interval-weights} have to be defined.

\begin{definition} [Adjustment of the expected interval-weights: Method 1]
Let the \textit{adjusted expected interval-weights} between vertices $i$ and $j$ be denoted as,
\begin{equation*}
E'^{I}=e'^{I}_{ij}=\Bigg[\frac{\underline{s}^{IO}_i \underline{s}^{IO}_j}{2\overline{w}'},\frac{\overline{s}^{IO}_i \overline{s}^{IO}_j}{2\underline{w}'}\Bigg],\qquad \left(0\notin [2\underline{w}',2\overline{w}']\right)
\end{equation*}
The adjustments for the minimum and maximum values are calculated as follows:
\setlength\itemsep{0.15cm}    
\begin{itemize}
	\item for $i=j$, the adjusted total weight, $2w'$, varies between
		\begin{equation}\label{chp5_adjust_i=j-min}
			2\underline{w}'=\Big[\underline{s}^{IO}_i,\underline{s}^{IO}_i\Big]+\sum\limits_{\substack{l=1\\ l\neq i}}^n s_l^{IO}
		\end{equation}	
		\begin{equation}\label{chp5_adjust_i=j-max}		
			2\overline{w}'=\Big[\overline{s}^{IO}_i,\overline{s}^{IO}_i\Big]+\sum\limits_{\substack{l=1\\ l\neq i}}^n s_l^{IO}
		\end{equation}
	Thus, when both limits of the interval $s_i^{IO}$ are at the \textit{minimum} value, the adjusted total weight is \textit{maximum} for $\max 2\underline{w}'$ (upper bound). Similarly, when both limits of the interval $s_i^{IO}$ are at the \textit{maximum} value, the adjusted total weight is \textit{minimum} for $\min 2\overline{w}'$ (lower bound). Then, the adjusted expected interval-weight when $i=j$ is denoted as:
	$$e'^{I}_{ij}=\Bigg[\frac{\underline{s}^{IO}_i \underline{s}^{IO}_j}{\max 2\underline{w}'},\frac{\overline{s}^{IO}_i \overline{s}^{IO}_j}{\min 2\overline{w}'}\Bigg],\qquad (0\notin [\min 2\overline{w}',\max 2\underline{w}'])$$
		
	\item for $i\neq j$, the adjusted total weight, $2w'$, varies between
		\begin{equation}\label{chp5_adjust_i-dif-j-min}
			2\underline{w}'=\Big[\underline{s}^{IO}_i,\underline{s}^{IO}_i\Big]+\Big[\underline{s}^{IO}_j,\underline{s}^{IO}_j\Big]+\sum\limits_{\substack{l=1\\ l\neq i\\ l\neq j}}^n s_l^{IO}
		\end{equation}
		\begin{equation}\label{chp5_adjust_i-dif-j-max}
			2\overline{w}'=\Big[\overline{s}^{IO}_i,\overline{s}^{IO}_i\Big]+\Big[\overline{s}^{IO}_j,\overline{s}^{IO}_j\Big]+\sum\limits_{\substack{l=1\\ l\neq i\\ l\neq j}}^n s_l^{IO}
		\end{equation}
	Likewise, when both limits of the interval $s_i^{IO}$ are at the \textit{minimum} value, the adjusted total weight is \textit{maximum} for $\max 2\underline{w}'$ (upper bound). Similarly, when both limits of the interval $s_i^{IO}$ are at the \textit{maximum} value, the adjusted total weight is \textit{minimum} for $\min 2\overline{w}'$ (lower bound). Then, the adjusted expected interval-weight when $i\neq j$ is denoted as:
	$$e'^{I}_{ij}=\Bigg[\frac{\underline{s}^{IO}_i \underline{s}^{IO}_j}{\max 2\underline{w}'},\frac{\overline{s}^{IO}_i \overline{s}^{IO}_j}{\min 2\overline{w}'}\Bigg],\qquad \left(0\notin [\min 2\overline{w}',\max 2\underline{w}']\right).$$
\end{itemize}		
\end{definition}

In the example that follows (Example~\ref{chp5_Ex:interval_tables_adjust-sum}), the \textit{expected interval-weighted contingency table} already takes into account the respective adjustments.

\begin{example}\label{chp5_Ex:interval_tables_adjust-sum}
Consider the same interval-weighted network from Example~\ref{chp5_Ex:interval_tables}. The adjusted contingency table for the expected interval-weights, $E^I$, can be written as follows:
\definecolor{Gray}{gray}{0.85}
\begin{figure}[H]
       \centering
    	\begin{subfigure}[c]{0.99\linewidth}
				\centering        
				\begin{tikzpicture}[inner sep=0pt, minimum size=5mm, auto]
   					\node[node_style,font=\fontsize{9}{10}\selectfont] (v1) at (0,1) {$v_1$};
    				\node[node_style,font=\fontsize{9}{10}\selectfont] (v2) at (1,2) {$v_2$};
	    			\node[node_style,font=\fontsize{9}{10}\selectfont] (v3) at (1,0) {$v_3$};
    				\node[node_style,font=\fontsize{9}{10}\selectfont] (v4) at (2.35,0) {$v_4$};
	    			\draw[edge_style]  (v1) edge node[above,sloped,pos=0.5,font=\fontsize{7}{10}\selectfont] {$[1,3]$} (v2);
    				\draw[edge_style]  (v1) edge node[below,sloped,pos=0.5,font=\fontsize{7}{10}\selectfont] {$[1,1]$} (v3);
    				\draw[edge_style]  (v2) edge node[right=0.1,pos=0.5,font=\fontsize{7}{10}\selectfont] {$[1,1]$} (v3);
    				\draw[edge_style]  (v3) edge node[below,pos=0.5,font=\fontsize{7}{10}\selectfont] {$[2,4]$} (v4);
				\end{tikzpicture}
        		\caption{} \label{chp5_fig2:interval-net_adjust}
        	\end{subfigure}%
        	             	
    	\begin{subfigure}[t]{0.49\linewidth}
			\setlength{\tabcolsep}{4pt}
			\setlength{\extrarowheight}{4pt}						
				\fontsize{8}{10}\selectfont
				\begin{tabular}{ cc| c c c c|c }
				&\multicolumn{6}{ c }{\footnotesize \textup{Observed interval-weights: $O^I=$}} \\
				 & & $v_1$ & $v_2$ & $v_3$ & $v_4$ & $s_i^{IO}$ \\ \hhline{~*{6}{-}}
				\multirow{4}{*}{}
 					& $v_1$ & \cellcolor{Gray}$[0,0]$ & $[1,3]$ & $[1,1]$ & $[0,0]$ & $[2,4]$ \\
					& $v_2$ & $[1,3]$ & \cellcolor{Gray}$[0,0]$ & $[1,1]$ & $[0,0]$ & $[2,4]$ \\  	
 					& $v_3$ & $[1,1]$ & $[1,1]$ & \cellcolor{Gray}$[0,0]$ & $[2,4]$ & $[4,6]$ \\ 
 					& $v_4$ & $[0,0]$ & $[0,0]$ & $[2,4]$ & \cellcolor{Gray}$[0,0]$ & $[2,4]$ \\ \cline{2-7}
 					& $s_j^{IO}$ & $[2,4]$ & $[2,4]$ & $[4,6]$ & $[2,4]$ & $[10,18]$ \\
				\end{tabular}
				\normalsize
				\caption{} \label{chp5_tab_O:Q-gain_1_adjust}
		\end{subfigure}%
		\begin{subfigure}[t]{0.49\linewidth}
			\setlength{\tabcolsep}{4pt}
			\setlength{\extrarowheight}{4pt}		
				\centering
				\fontsize{8}{10}\selectfont
				\begin{tabular}{ cc| c c c c|c }
				&\multicolumn{6}{ c }{\footnotesize \textup{Adjusted Expected interval-weights: $E^I=$}} \\
				 & & $v_1$ & $v_2$ & $v_3$ & $v_4$ &  \\ \hhline{~*{6}{-}}
				\multirow{4}{*}{}
 					& $v_1$ & \cellcolor{Gray}$\Big[\frac{4}{16},\frac{16}{12}\Big]$ & $\Big[\frac{4}{14},\frac{16}{14}\Big]$ & $\Big[\frac{8}{14},\frac{24}{14}\Big]$ & $\Big[\frac{4}{14},\frac{16}{14}\Big]$ & $$\\
					& $v_2$ & $\Big[\frac{4}{14},\frac{16}{14}\Big]$ & \cellcolor{Gray}$\Big[\frac{4}{16},\frac{16}{12}\Big]$ & $\Big[\frac{8}{14},\frac{24}{14}\Big]$ & $\Big[\frac{4}{14},\frac{16}{14}\Big]$ & $$ \\  	
 					& $v_3$ & $\Big[\frac{8}{14},\frac{24}{14}\Big]$ & $\Big[\frac{8}{14},\frac{24}{14}\Big]$ & \cellcolor{Gray}$\Big[\frac{16}{16},\frac{36}{12}\Big]$ & $\Big[\frac{8}{14},\frac{24}{14}\Big]$ & $$\\ 
 					& $v_4$ & $\Big[\frac{4}{14},\frac{16}{14}\Big]$ & $\Big[\frac{4}{14},\frac{16}{14}\Big]$ & $\Big[\frac{8}{14},\frac{24}{14}\Big]$ & \cellcolor{Gray}$\Big[\frac{4}{16},\frac{16}{12}\Big]$ & $$\\ \cline{2-7}
 					&  &  &  & &  &  \\
				\end{tabular}
				\normalsize
				\caption{} \label{chp5_tab_E:Q-gain_1_adjust}
		\end{subfigure}	%

\vspace{1cm}
	\begin{subfigure}[t]{0.99\linewidth}
		\fontsize{8}{10}\selectfont
		\centering
		\begin{adjustbox}{max width=\textwidth}	
			\renewcommand{\arraystretch}{0.9}
			\begin{tabular}{l|l|l}	
				\hline
				& \multicolumn{2}{c}{\scriptsize \textup{Total weight adjustments}} \\ \cline{2-3}
				\scriptsize \textup{Vertices} & \multicolumn{1}{c|}{\scriptsize \textup{Adjusted minimum}} & \multicolumn{1}{c}{\scriptsize \textup{Adjusted maximum}}
				\\ \hline
				$v_1 - v_1$ & $4+2+4+2=12$ & $2+4+6+4=16$ \\\hline
				$v_1 - v_2$ & $4+4+4+2=14$ & $2+2+6+4=14$ \\\hline
				$v_1 - v_3$ & $4+2+6+2=14$ & $2+4+4+4=14$ \\\hline
				$v_1 - v_4$ & $4+2+4+4=14$ & $2+4+6+2=14$ \\\hline
				$v_2 - v_2$ & $2+4+4+2=12$ & $4+2+6+4=16$ \\\hline
				$v_2 - v_3$ & $2+4+6+2=14$ & $4+2+4+4=14$ \\\hline
				$v_2 - v_4$ & $2+4+4+4=14$ & $4+2+6+2=14$ \\\hline
				$v_3 - v_3$ & $2+2+6+2=12$ & $4+4+4+4=16$ \\\hline
				$v_3 - v_4$ & $2+2+6+4=14$ & $4+4+4+2=14$ \\\hline
				$v_4 - v_4$ & $2+2+4+4=12$ & $4+4+6+2=16$ \\\hline
			\end{tabular}
		\end{adjustbox}
		\caption{}
		\label{chp5_tab:Q-gain_adjust}
	\end{subfigure}%
\caption[Example of the \textit{adjustments} for the expected interval-weights (Method 1)]{(a) Interval-weighted network - $G^I$, (b) tabular representation of the \textit{observed interval-weights} -- $O^I$, (c) tabular representation of the adjusted \textit{expected interval-weights} -- $E^I$ and, (d) Adjustments for the total interval weights.}
\end{figure}
\end{example}

To exemplify how the values in Table~\ref{chp5_tab:Q-gain_adjust} of Example~\ref{chp5_Ex:interval_tables_adjust-sum} were obtained both for cases $(i=j)$ (expressions~(\ref{chp5_adjust_i=j-min}) and~(\ref{chp5_adjust_i=j-max})) and $(i\neq j)$ (expressions~(\ref{chp5_adjust_i-dif-j-min}) and~(\ref{chp5_adjust_i-dif-j-max})), we detail below the calculations for the pairs of vertices $(v_1,v_1)$ and $(v_1,v_2)$:

\begin{itemize}
\setlength\itemsep{0.15cm} 
\item vertices $(v_1,v_1)$:
	{\small	
	\begin{itemize}[leftmargin=0.5cm]
	\item [-]\textit{adjusted minimum} $=\min\big\{2\overline{w}'\big\}=\min\big\{[4,4]{+}[2,4]{+}[4,6]{+}[2,4]\big\}=\min\big\{[12,18]\big\}=12$;
	\item [-]\textit{adjusted maximum} $=\max\big\{2\underline{w}'\big\}=\max\big\{[2,2]{+}[2,4]{+}[4,6]{+}[2,4]\big\}=\max\big\{[10,16]\big\}=16$;
	\end{itemize}
	}%
\item vertices $(v_1,v_2)$:
	{\small		
	\begin{itemize}[leftmargin=0.5cm]
	\item [-] \textit{adjusted minimum} $=\min\big\{2\overline{w}'\big\}=\min\big\{[4,4]{+}[4,4]{+}[4,6]{+}[2,4]\big\}=\min\big\{[14,18]\big\}=14$;
	\item [-] \textit{adjusted maximum} $=\max\big\{2\underline{w}'\big\}=\max\big\{[2,2]{+}[2,2]{+}[4,6]{+}[2,4]\big\}=\max\big\{[10,14]\big\}=14$.\\
	\end{itemize}
	}%
\end{itemize}

The pseudo-code of the ``Method 1. Classic Louvain'' algorithm is presented below in Algorithm~\ref{chp4_alg:Classic_Louvain_1}.

\begin{algorithm}[H]
\caption{Pseudo-code: ``Method 1. Classic Louvain'' algorithm for IWN}\label{chp4_alg:Classic_Louvain_1}
\textbf{Input:} An interval-weighted network $G^I=(V^I,E^I,W^I)$\\
\textbf{Output:} A partition of $G^I$ into communities 
\begin{enumerate}[itemsep=0.5pt,label*=\arabic*:]
\item Initialization: \textit{each vertex forms a community}
\item \textit{Phase 1}: Modularity optimization using \textbf{\textit{intervals}} (refine communities)\tikzmark{top}
   \item \qquad Repeat iteratively for all vertices $i$
                \item \qquad \qquad Remove $i$ from its community
                    \item \qquad \qquad \qquad Compute $\Delta Q^I_{i\to C_j}$ for each neighbour $j$
                \item \qquad \qquad Insert $i$ in a neighbouring community of $i$ so as to maximize modularity
                    \item \qquad \qquad \qquad Join the community $C_j$ that yields the largest gain in modularity $\Delta Q^I$\tikzmark{right}
     \item \qquad  Repeat until no movement yields a gain in modularity
\item \textit{Phase 2:} Community aggregation (reconstruct the network)
   \item \qquad The communities become \textit{super-vertices}
   \item \qquad The \textit{intervals} on the edges between the formed communities are \textbf{\textit{summed}}\tikzmark{bottom}
\item Repeat steps (2) to (9) until convergence (stop when the modularity cannot be increased)
\end{enumerate}
\newcommand*{\AddNote}[4]{%
    \begin{tikzpicture}[overlay, remember picture]
        \draw [decoration={brace,amplitude=0.5em}, decorate, thick, darkgray]
            ([xshift=2mm]$(#3)!(#1.north)!($(#3)+(0,-1)$)$) --  
            ([xshift=2mm]$(#3)!(#2.south)!($(#3)+(0,-1)$)$)
                node [align=center, text width=2cm, pos=0.5, anchor=east, xshift=2cm] {#4};
    \end{tikzpicture}
}%
\AddNote{top}{bottom}{right}{\textcolor{darkgray}PASS\break \vspace{0.01mm} {\scriptsize Phase 1\break +\break Phase 2}}
\end{algorithm}

Finally, in Table~\ref{chp5_tab_Method 1:intervals sum} depicted below, are the results for the ``Method 1. Intervals sum'' (1\textsuperscript{st} phase = Sum and 2\textsuperscript{nd} phase = Sum) of the Louvain algorithm for interval-weighted networks corresponding to the 1\textsuperscript{st} iteration of the 1\textsuperscript{st} pass. This method detected the aggregation of the four vertices in two communities, $C_1=\{v_1,v_2\}$, and $C_2=\{v_3,v_4\}$.
In Appendix~\ref{appendix_C}, the complete Louvain algorithm output for all generated steps that led to the results in Table~\ref{chp5_tab_Method 1:intervals sum} is shown.\\

\newpage

\begin{table}[H]
\centering
\begin{threeparttable}
\caption{Modularity gain results for the 1\textsuperscript{st} iteration of the 1\textsuperscript{st} pass of the Louvain algorithm for interval-weighted networks (Method 1. Intervals sum: Phase 1 = Sum / Phase 2 = Sum.}
\label{chp5_tab_Method 1:intervals sum}
\fontsize{8}{10}\selectfont
\renewcommand{\arraystretch}{1.1}
\begin{tabular}{c?c?c}
\thickhline
\multicolumn{2}{c?}{\multirow{2}{*}{}} & \multicolumn{1}{c}{\cellcolor{gray!20} Method 1: Intervals Sum}                                                 \\ \cline{3-3}
\multicolumn{2}{c?}{} & \multicolumn{1}{c}{{\scriptsize{Difference -- $D$ \tnote{a}}}}\\ \cline{3-3} 
\multicolumn{2}{c?}{} & \multicolumn{1}{c}{{\scriptsize{Interval Modularity--1: $Q^I=\sum_r D\left(o_{rr},e_{rr}\right)$}}} \\ \cline{3-3}
\multicolumn{2}{l?}{Vertices} & {\scriptsize{Modularity gain for IWN: $\Delta Q^I=Q^I_{new} - Q^I_{last}$}} \\ \thickhline
\multirow{2}{*}{$v_1$} & $\mathbf{v_1\to v_2}$ 	&	$\mathbf{\hphantom{-}4.095}$	\\ \cline{2-3}
                       & $v_1\to v_3$ 			&	$-0.810$						\\ \thickhline
\multirow{2}{*}{$v_2$} & $v_2\to v_1,v_2$			&	$\hphantom{-}4.095$			\\ \cline{2-3}
                       & $v_2\to v_3$ 			&	$-0.810$						\\ \thickhline
\multirow{2}{*}{$v_3$} & $v_3\to v_1,v_2$			&	$-0.679$						\\ \cline{2-3}
                       & $\mathbf{v_3\to v_4}$ 	&	$\mathbf{\hphantom{-}5.762}$	\\ \thickhline
$v_4$					&	$v_4\to v_3,v_4$ 	 	&	$\hphantom{-}5.762$			\\ \thickhline
$\vdots$ 				& $\vdots$ 					&	$\vdots$ 						\\ \thickhline
\multicolumn{2}{c?}{{\scriptsize{No. final communities}}}	&	$2$					\\ \thickhline
\end{tabular}%
\begin{tablenotes}
      \tiny
		\item{The values that led to the movement from one vertex to another community are highlighted in bold.}
		\item[a]{$D\big([\underline{x},\overline{x}],[\underline{y},\overline{y}]\big)=\max\big\{|\underline{x}-\underline{y}|,|\overline{x}-\overline{y}|\big\}\times sign\ argmax\big\{|\underline{x}-\underline{y}|,|\overline{x}-\overline{y}|\big\}$.}
\end{tablenotes}
\end{threeparttable}
\end{table}

\subsection{Hybrid Louvain -- Method 2: Intervals Midpoint}
\label{chp5_subsction_CL--Method 2: Intervals Min--Max}

The second method we developed to detect communities in interval-weighted networks, based on the Louvain algorithm, is characterized by the following -- for each \textit{pass} of the Louvain algorithm, on the 1\textsuperscript{st} phase (\textit{the optimization phase}), modularity and modularity gain are computed by \textit{summing the midpoints} of the intervals (identical to what is done when considering a weighted network, see Section~\ref{chp5:sec_1}); the 2\textsuperscript{nd} phase (\textit{the aggregation of the network}) is done by selecting the \textit{minimum} and the \textit{maximum} values of the intervals for the formed communities (Definition~\ref{Def_supervertices_HybridLouvain}). 

\begin{definition}\label{Def_supervertices_HybridLouvain}
Let us denote by $C_1$ and $C_2$ two communities in the original interval-weighted network $(G^I)$, where $O^I{=}o^I_{ij}{=}\Big[\underline{o}_{ij},\overline{o}_{ij}\Big]$. When creating the ``super-vertices'' in the aggregated IWN $G^{I'}=(V^{I'},E^{I'},W^{I'})$, the interval-valued weight of an edge is defined as follows:
\begin{equation}\label{chp4_Eq:phase2-super_edges}
o^{I'}_{C_1C_2}=\left[\min\limits_{\substack{i\in C_1\\ j\in C_2}}\{\underline{o}_{ij}\},\max\limits_{\substack{i\in C_1\\ j\in C_2}}\{\overline{o}_{ij}\}\right]
\end{equation}
\end{definition}

The pseudo-code of the Hybrid Louvain algorithm is presented in Algorithm~\ref{chp4_alg:Hybrid_Louvain_1}. Not using intervals in the modularity optimization phase calculations, this method revealed computationally less expensive than the previous one.

\begin{algorithm}[H]
\caption{Pseudo-code: ``Method 2. Hybrid Louvain'' algorithm for IWN}\label{chp4_alg:Hybrid_Louvain_1}
\textbf{Input:} An interval-weighted network $G^I=(V^I,E^I,W^I)$\\
\textbf{Output:} A partition of $G^I$ into communities 
\begin{enumerate}[itemsep=0.5pt,label*=\arabic*:]
\item Initialization: \textit{each vertex forms a community}
\item \textit{Phase 1}: Modularity optimization using \textbf{\textit{intervals midpoints}} (refine communities)\tikzmark{top}
   \item \qquad Repeat iteratively for all vertices $i$
                \item \qquad \qquad Remove $i$ from its community
                    \item \qquad \qquad \qquad Compute $\Delta Q^I_{i\to C_j}$ for each neighbour $j$
                \item \qquad \qquad Insert $i$ in a neighbouring community of $i$ so as to maximize modularity
                    \item \qquad \qquad \qquad Join the community $C_j$ that yields the largest gain in modularity $\Delta Q^I$\tikzmark{right}
     \item \qquad  Repeat until no movement yields a gain in modularity
\item \textit{Phase 2:} Community aggregation (reconstruct the network)
   \item \qquad The communities become \textit{super-vertices}
   \item \qquad The weights of the edges between communities are the \textbf{\textit{minimum}} and the\\
   				\textbf{\textit{maximum}} values of the intervals for the formed communities\tikzmark{bottom}
\item Repeat steps (2) to (9) until convergence (stop when the modularity cannot be increased)
\end{enumerate}
\newcommand*{\AddNote}[4]{%
    \begin{tikzpicture}[overlay, remember picture]
        \draw [decoration={brace,amplitude=0.5em}, decorate, thick, darkgray]
            ([xshift=2mm]$(#3)!(#1.north)!($(#3)+(0,-1)$)$) --  
            ([xshift=2mm]$(#3)!(#2.south)!($(#3)+(0,-1)$)$)
                node [align=center, text width=2cm, pos=0.5, anchor=east, xshift=2cm] {#4};
    \end{tikzpicture}
}%
\AddNote{top}{bottom}{right}{\textcolor{darkgray}PASS\break \vspace{0.01mm} {\scriptsize Phase 1\break +\break Phase 2}}
\end{algorithm}

\noindent
\textbf{Creating the new interval-weighted network (coarsening the network) at Louvain's algorithm Phase 2}

In the classic Louvain algorithm~\citep{Blondel:2008do}, the 2\textsuperscript{nd} Phase consists in building a new network, whose vertices are the communities found in the previous iteration (1\textsuperscript{st} Phase). The input network is collapsed, and the weights of the edges between the new ``super-vertices'' are given by the \textit{sum} of the weights between all the vertices in the old communities. This imposes that the creation of a community of ``super-vertices'' in the aggregated network should be equivalent to clustering all the vertices of the associated communities in the original network.

\begin{example}\label{Example_supervertices_Louvain}
Let us denote by $C_1$ and $C_2$ two communities in the original weighted network $G^W(V,E,W)$ which become ``super-vertices'' in the aggregated network $G^{W'}(V',E',W')$. Then, for the modularity, one needs to impose that:
\begin{equation}\label{chp4_Eq:phase2-super_vertices}
\sum\limits_{\substack{i\in C_1\\ j\in C_2}} \bigg[w_{ij}-\frac{s_i s_j}{2w}\bigg]=w'_{C_1C_2}-\frac{s'_{C_1} s'_{C_2}}{2w'}
\end{equation}
which leads to defining an edge between two vertices in the aggregated network as the sum of the edges between the two associated communities in the original network,
\begin{equation}\label{chp4_Eq:phase2-super_edges}
w'_{C_1C_2}=\sum\limits_{\substack{i\in C_1\\ j\in C_2}} w_{ij}.
\end{equation}

The same is true for the total strength,
\begin{align}
2w'&=\sum_{C_1,C_2\in V'} w'_{C_1 C_2}=\sum_{C_1,C_2\in V'} \sum\limits_{\substack{i\in C_1\\ j\in C_2}} w{ij}\\\nonumber
		&=\sum_{C_1\in V'} \sum_{i\in C_1} \sum_{C_2\in V'} \sum_{j\in C_2} w_{ij}=\sum_{i\in V}\sum_{j\in V} w_{ij}=2w
\end{align}
\noindent
this implies that~(\ref{chp4_Eq:phase2-super_vertices}) is satisfied.
\end{example}

Consider the interval-weighted network of Figure~\ref{chp5_Fig:HL_Method2.1_a}. First the algorithm calculates the intervals' midpoints of the network's edges (Figure~\ref{chp5_Fig:HL_Method2.1_b}) and only then applies the optimization phase of Louvain's algorithm using these values (Tables~\ref{chp5_Tab:HL_Method2.1_O} and~\ref{chp5_Tab:HL_Method2.1_E}) for the modularity gain calculations (Phase 1 of the Louvain algorithm).

\definecolor{Gray}{gray}{0.85}
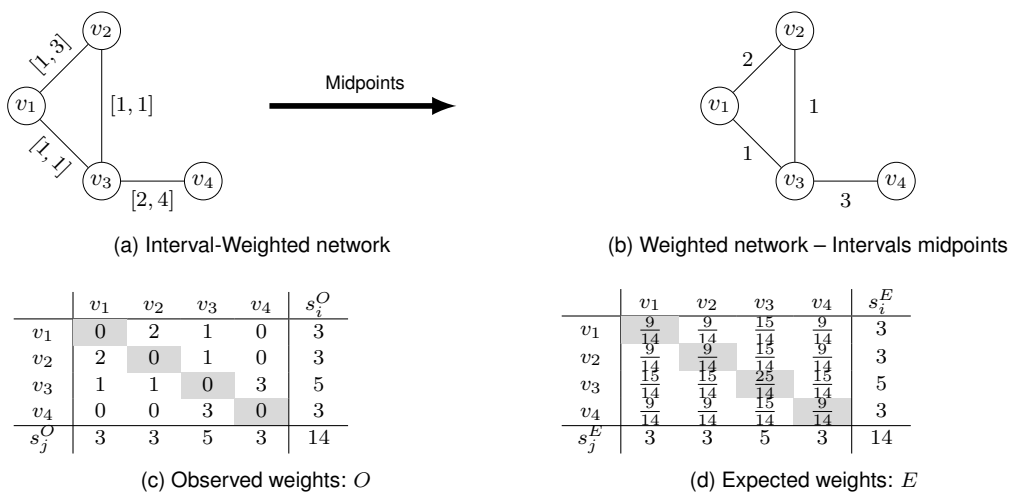
\begin{figure}[ht]
       \centering
    	\begin{subfigure}[c]{0.45\linewidth}
				\centering        
				\begin{tikzpicture}[inner sep=0pt, minimum size=5mm, auto]
   					\node[node_style,font=\fontsize{9}{10}\selectfont] (v1) at (0,1) {$v_1$};
    				\node[node_style,font=\fontsize{9}{10}\selectfont] (v2) at (1,2) {$v_2$};
	    			\node[node_style,font=\fontsize{9}{10}\selectfont] (v3) at (1,0) {$v_3$};
    				\node[node_style,font=\fontsize{9}{10}\selectfont] (v4) at (2.35,0) {$v_4$};
	    			\draw[edge_style]  (v1) edge node[above,sloped,pos=0.5,font=\fontsize{8}{10}\selectfont] {$[1,3]$} (v2);
    				\draw[edge_style]  (v1) edge node[below,sloped,pos=0.5,font=\fontsize{8}{10}\selectfont] {$[1,1]$} (v3);
    				\draw[edge_style]  (v2) edge node[right=0.1,pos=0.5,font=\fontsize{8}{10}\selectfont] {$[1,1]$} (v3);
    				\draw[edge_style]  (v3) edge node[below,pos=0.5,font=\fontsize{8}{10}\selectfont] {$[2,4]$} (v4);
\hspace{0.5cm}
	\node[] (v*) at (3,1) {};
	\node[] (v**) at (6,1) {};  
	\draw[-latex,line width=0.75mm, above,pos=0.3,font=\fontsize{7}{10}\selectfont, shorten <= -15pt, shorten >= 15pt] (v*) -- node [] {Midpoints} (v**);    				
				\end{tikzpicture}
				\caption{Interval-Weighted network}\label{chp5_Fig:HL_Method2.1_a}
			\end{subfigure}    	
			\begin{subfigure}[c]{0.45\linewidth}
				\centering        
        		\begin{tikzpicture}[inner sep=0pt, minimum size=5mm, auto]
   					\node[node_style,font=\fontsize{9}{10}\selectfont] (v1) at (0,1) {$v_1$};
    				\node[node_style,font=\fontsize{9}{10}\selectfont] (v2) at (1,2) {$v_2$};
	    			\node[node_style,font=\fontsize{9}{10}\selectfont] (v3) at (1,0) {$v_3$};
    				\node[node_style,font=\fontsize{9}{10}\selectfont] (v4) at (2.35,0) {$v_4$};
	    			\draw[edge_style]  (v1) edge node[above,pos=0.3,font=\fontsize{8}{10}\selectfont] {$2$} (v2);
    				\draw[edge_style]  (v1) edge node[below,pos=0.3,font=\fontsize{8}{10}\selectfont] {$1$} (v3);
    				\draw[edge_style]  (v2) edge node[right,pos=0.5,font=\fontsize{8}{10}\selectfont] {$1$} (v3);
    				\draw[edge_style]  (v3) edge node[below,pos=0.5,font=\fontsize{8}{10}\selectfont] {$3$} (v4);
				\end{tikzpicture}
        		\caption{Weighted network -- Intervals midpoints}\label{chp5_Fig:HL_Method2.1_b}
        	\end{subfigure}%
        	             	
    	\begin{subfigure}[b]{0.45\linewidth}
				\fontsize{8}{10}\selectfont
				\begin{tabular}{ cc| c c c c|c }
				&\multicolumn{6}{ c }{} \\
				 & & $v_1$ & $v_2$ & $v_3$ & $v_4$ & $s_i^O$ \\ \hhline{~*{6}{-}}
				\multirow{4}{*}{}
 					& $v_1$ & \cellcolor{Gray}$0$ & $2$ & $1$& $0$ & $3$ \\
					& $v_2$ & $2$ & \cellcolor{Gray}$0$ & $1$ & $0$ & $3$ \\  	
 					& $v_3$ & $1$ & $1$ & \cellcolor{Gray}$0$ & $3$ & $5$ \\ 
 					& $v_4$ & $0$ & $0$ & $3$ & \cellcolor{Gray}$0$ & $3$ \\ \cline{2-7}
 					& $s_j^O$ & $3$ & $3$ & $5$ & $3$ & $14$ \\
				\end{tabular}
				\caption{Observed weights: $O$}\label{chp5_Tab:HL_Method2.1_O}
		\end{subfigure}%
		\begin{subfigure}[b]{0.45\linewidth}
				\fontsize{8}{10}\selectfont
				\begin{tabular}{ cc| c c c c|c }
				&\multicolumn{6}{ c }{} \\
				 & & $v_1$ & $v_2$ & $v_3$ & $v_4$ & $s_i^E$ \\ \hhline{~*{6}{-}}
				\multirow{4}{*}{}
 					& $v_1$ & \cellcolor{Gray}$\frac{9}{14}$ & $\frac{9}{14}$ & $\frac{15}{14}$ & $\frac{9}{14}$ & $3$ \\
					& $v_2$ & $\frac{9}{14}$ & \cellcolor{Gray}$\frac{9}{14}$ & $\frac{15}{14}$ & $\frac{9}{14}$ & $3$ \\  	
 					& $v_3$ & $\frac{15}{14}$ & $\frac{15}{14}$ & \cellcolor{Gray}$\frac{25}{14}$ & $\frac{15}{14}$ & $5$\\ 
 					& $v_4$ & $\frac{9}{14}$ & $\frac{9}{14}$ & $\frac{15}{14}$ & \cellcolor{Gray}$\frac{9}{14}$ & $3$\\ \cline{2-7}
 					& $s_j^E$ & $3$ & $3$ & $5$ & $3$ & $14$ \\
				\end{tabular}
				\caption{Expected weights: $E$}\label{chp5_Tab:HL_Method2.1_E}
		\end{subfigure}	%
\caption[Hybrid Louvain -- Method 2: Intervals midpoint]{Hybrid Louvain -- Method 2: Intervals midpoints.}
\label{chp5_Fig:HL_Method2.1}
\end{figure}

\begin{figure}[H]
	\centering
    \includegraphics[scale=0.45, clip, trim={0cm 1cm 0cm 1cm}]{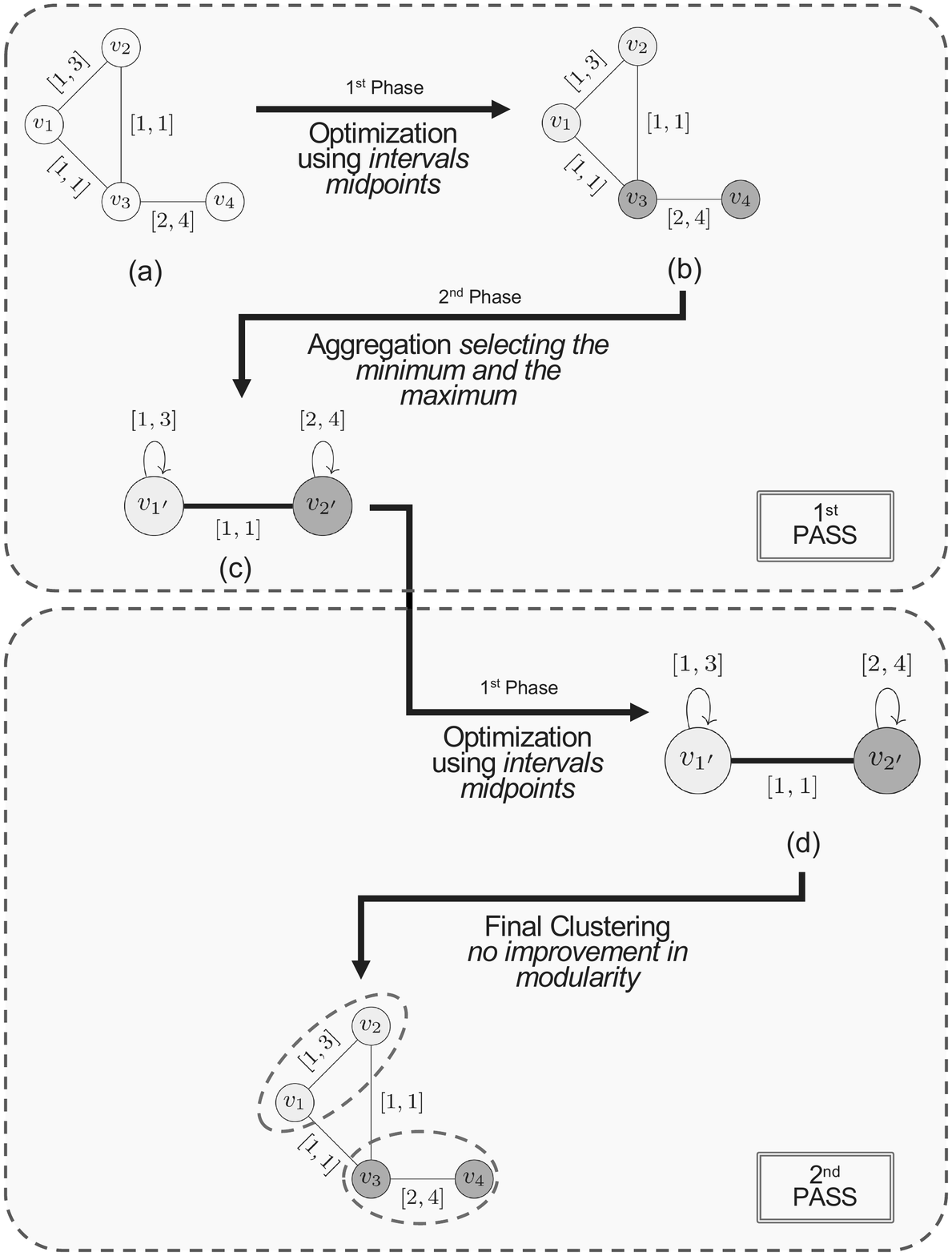}
     \caption[Sketch of the optimization and aggregation steps of the Hybrid Louvain algorithm for IWN]{Sketch of the optimization and aggregation steps of the Hybrid Louvain algorithm for IWN. (a) $1^{st}$ Pass: Initial network corresponds to the initial partition (4 communities: $Q=-3.714$), (b) $1^{st}$ Pass: local optimization (2 communities: $Q=2.857$), (c) $1^{st}$ Pass: network aggregation (2 communities: $Q=2.857$), (d) $2^{nd}$ Pass: local optimization (2 communities: $Q=2.857$) and Final clustering (modularity has reached its maximum).}
     \label{chp4_fig:Hybrid_Louvain-method}
\end{figure}

This process is called an \textit{iteration}, and is applied sequentially to all the vertices. The process is then repeated for all the vertices until no further improvements are obtained in a complete iteration, i.e., when the modularity has reached a local optimum, which implies that no vertex migration increases the modularity. The first phase is then finished. The subsequent step of the algorithm starts (Phase 2), consisting in building a new network, whose vertices are the communities found in the previous iteration (Phase 1). The input interval-weighted network is collapsed, and the intervals associated with the edges between the new ``super''-vertices are given by the \textit{minimum} and the \textit{maximum} of the intervals between all the vertices in the old communities. Likewise, the edges and vertices within a community lead to \textit{loops} in the new network, weighted by the \textit{minimum} and the \textit{maximum} edge weights between the included vertices (see Figure~\ref{chp4_fig:Hybrid_Louvain-method}).

After completing the second phase, the algorithm completes one ``pass'' and goes back to the first phase in order to make multiple passes. This iterative procedure produces one partition per pass, thus creating a hierarchy of communities. The algorithm repeats these passes iteratively until the communities become stable, that is, until a maximum of modularity is reached, as depicted in Figure~\ref{chp4_fig:Hybrid_Louvain-method}.\\

Table~\ref{chp5_tab_Method 2.1:intervals midpoint} shows the results for the modularity gain calculations for the 1\textsuperscript{st} iteration of the 1\textsuperscript{st} pass of the Louvain algorithm for this method. In Appendix~\ref{appendix_A}, the complete Louvain algorithm output for all generated steps that led to the results in Table~\ref{chp5_tab_Method 2.1:intervals midpoint} is shown.\\

\newpage

\begin{table}[H]
\centering
\begin{threeparttable}
\caption{Modularity gain results for the 1\textsuperscript{st} iteration of the 1\textsuperscript{st} pass of the Louvain algorithm for interval--weighted networks (Method 2. Intervals Midpoint: Phase 1 = Midpoints Sum / Phase 2 = Min-Max).}
\label{chp5_tab_Method 2.1:intervals midpoint}
\fontsize{8}{10}\selectfont
\renewcommand{\arraystretch}{1.1}
\begin{tabular}{M{0.5cm}?M{1.5cm}?M{6cm}}
\thickhline
\multicolumn{2}{c?}{\multirow{2}{*}{}} & \cellcolor{gray!20} Method 2. Intervals Midpoint\\ \cline{3-3}
\multicolumn{2}{c?}{} & \multicolumn{1}{c}{{\scriptsize{Difference -- $D$ \tnote{a}}}}\\ \cline{3-3} 
\multicolumn{2}{c?}{} & \multicolumn{1}{c}{{\scriptsize{Interval Modularity: $Q^I=\sum_r D\left(o_{rr},e_{rr}\right)$}}} \\ \cline{3-3}
\multicolumn{2}{l?}{Vertices} & {\scriptsize{Modularity gain for IWN: $\Delta Q^I=Q^I_{new} - Q^I_{last}$}} \\ \thickhline
\multirow{2}{*}{$v_1$} &	$\mathbf{v_1\to v_2}$ 	 & $\hphantom{-}\mathbf{2.714}$	\\ \cline{2-3}
                       & $v_1\to v_3$ 	&	$-0.143$  \\ \thickhline
\multirow{2}{*}{$v_2$} &	$v_2\to v_1,v_2$  & $\hphantom{-}2.714$ \\ \cline{2-3}
                       & $v_2\to v_3$ 	& $-0.143$ \\ \thickhline
\multirow{2}{*}{$v_3$} & $v_3\to v_1,v_2$ & $-0.284$\\ \cline{2-3}
                       &	$\mathbf{v_3\to v_4}$ 	 & $\hphantom{-}\mathbf{3.857}$\\ \thickhline
\multirow{1}{*}{$v_4$} &	$v_4\to v_3,v_4$ 	& $\hphantom{-}3.857$ \\ \hline
$\vdots$ & $\vdots$ & $\vdots$ \\ \thickhline
\multicolumn{2}{c?}{No. final Communities} & $2$\\ \thickhline
\end{tabular}%
\begin{tablenotes}
      \tiny
		\item{The values that led to the movement from one vertex to another community are highlighted in bold.}
		\item[a]{$D\big([\underline{x},\overline{x}],[\underline{y},\overline{y}]\big)=\max\big\{|\underline{x}-\underline{y}|,|\overline{x}-\overline{y}|\big\}\times sign\ argmax\big\{|\underline{x}-\underline{y}|,|\overline{x}-\overline{y}|\big\}$.}
\end{tablenotes}
\end{threeparttable}
\end{table}

\section{A real-world example: Portuguese commuters}
\label{example_commuters}

In recent years, community detection techniques and centrality measures have often been used in complex networks representing territorial units as tools to identify homogeneous groups of these units~\citep{DeMontis:2013ho,DeMontis:2013dl,Traag:2009et,Barigozzi:2011bd,Traag:2014tj}. We present the application of our community detection method to a real-world interval-weighted commuters network. In this network we analyse the community structure that emerges from the movements of daily commuters in mainland Portugal (by all means of transportation) between the twenty three NUTS 3 Regions (source: INE -- Statistics Portugal, Census 2011)\footnote{NUTS--Nomenclature of Territorial Units for Statistics~\citep{Eurostat:2016a}.} (henceforth, the ``Interval-Weighted Commuters Network (IWCN)''), through the application of each of the network community detection methods developed for Interval-Weighted Networks (IWN).
%

\begin{figure}[ht]
	\centering
    	\includegraphics[scale=0.8, clip, trim={10cm 7.5cm 9.5cm 4cm}]{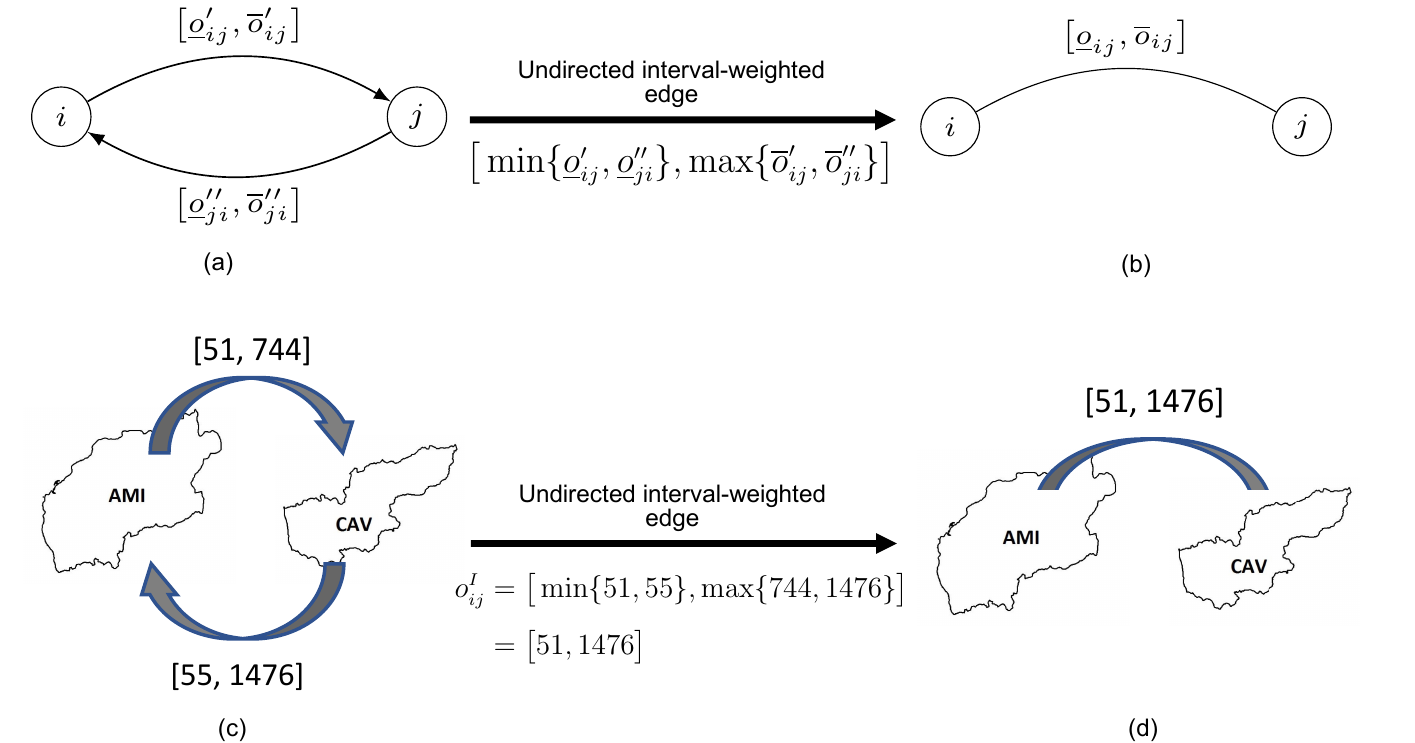}
	\caption[Sketch of the conversion of directed interval-weighted edges into an undirected interval--weighted edge]{Sketch of the conversion of directed interval-weighted edges into an undirected interval-weighted edge. (a) Bidirectional interval flows $i\to j$ and $j\to i$ between NUTS 3 $i$ and $j$, (b) Undirected interval flow between NUTS 3 $i$ and $j$, (c) and (d) are an example extracted from the real data, where NUTS 3 $i= \text{Alto Minho}\ \text{and}\ j= \text{Cavado}$.}
	\label{chp7_fig:directed_to_undirected_edges}
\end{figure}

Each \textit{vertex} of the Interval-Weighted Commuters Network (IWCN) corresponds to a given NUTS 3 (which in turn represents the aggregation of commuter flows between the municipalities that constitute that region) and the \textit{edges} are associated with intervals ranging between the \textit{minimum} flow \textit{larger than 50 commuters} and \textit{maximum} flow of commuters among the corresponding NUTS 3. As represented in Figure~\ref{chp7_fig:directed_to_undirected_edges}a, the interval of commuters flow from NUTS $i\to j$ may be  different from the one of $j\to i$. Therefore, the elements $o^I_{ij}$ of the symmetric interval-weighted adjacency matrix, $O^I$, denote the maximum variability of the \textit{bi-directional} flows $ij$ and $ji$ between the NUTS $i$ and $j$ (Figure\ref{chp7_fig:directed_to_undirected_edges}b): \linebreak$o^I_{ij}=\big[\min\{\underline{o}'_{ij},\underline{o}''_{ji}\},\max\{\overline{o}'_{ij},\overline{o}''_{ji}\}\big]=\big[\underline{o}_{ij},\overline{o}_{ij}\big]$. The option for this representation of flows is related to the fact that we do not want to study the direction of these daily commuter fluxes, but just quantify the reciprocal attractiveness of the NUTS 3 pairs~\citep{DeMontis:2013ho}. This kind of aggregation when the data are recorded at the same point in time and the statistical units to be analysed are not those for which the data was originally recorded, but constitute specific groups of those (higher level than the one at which the data was originally collected), is called \textit{contemporary aggregation}~\citep{Brito:2014es}.

The adjacency matrix elements are null, $o^I_{ij}=[0,0]$, when there is no commuter flow greater than 50 daily movements between NUTS 3 $i$ and $j$. By definition, we assume that there are no commuter flows within each NUTS 3, i.e., the network has no loops at initial  vertices, which implies that the diagonal of the interval-weighted adjacency matrix consists of degenerate intervals with the value zero, $o^I_{ii}=[0,0]$.

\begin{figure}[H]
    \centering
    \begin{subfigure}[t]{0.48\linewidth}
        \centering
        \includegraphics[width=0.5\linewidth,clip, trim={0cm 0cm 0cm 0cm}]{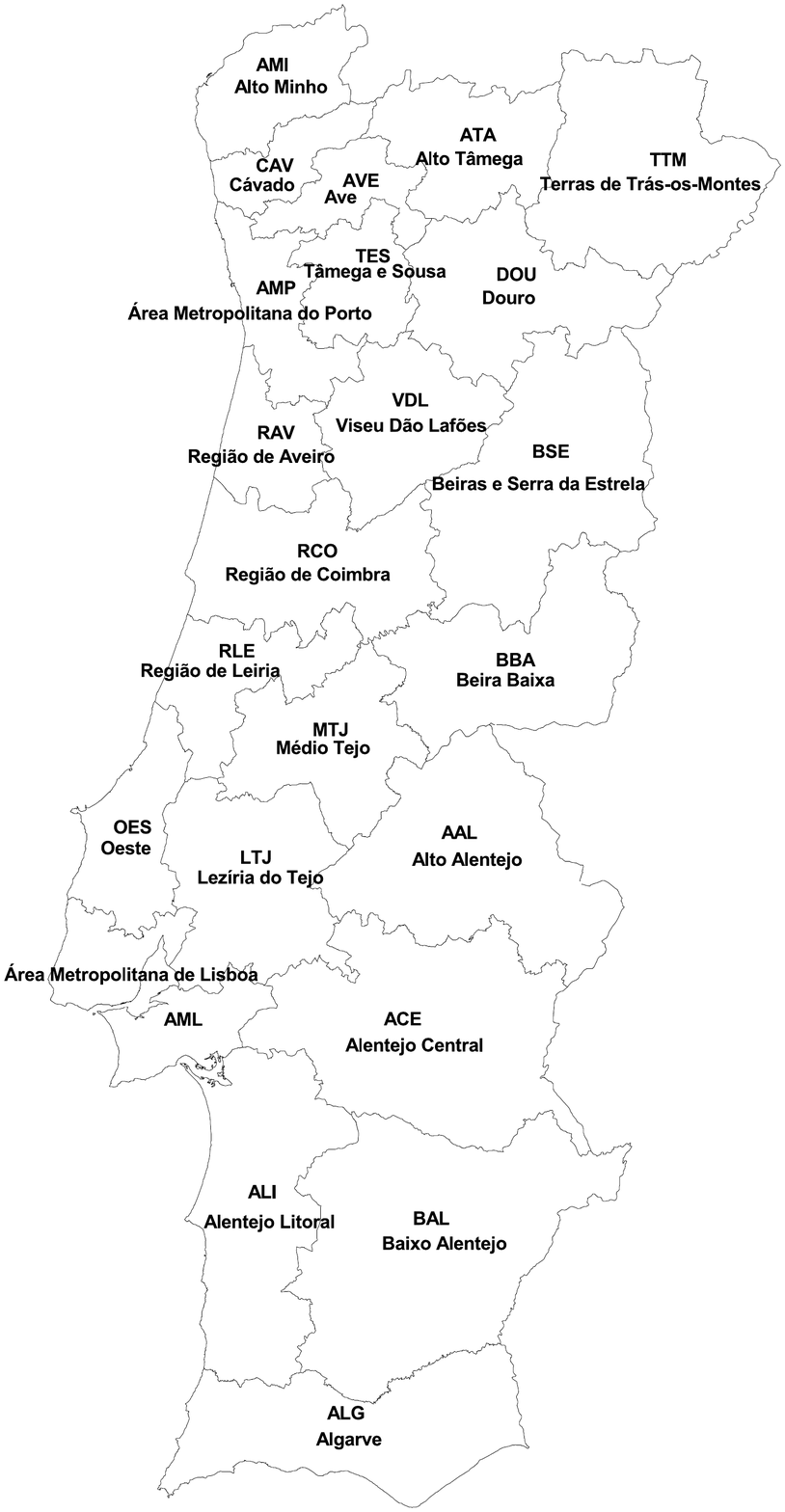}
        \caption{}
        \label{chp7_fig:Map_PT}
    \end{subfigure}
    \begin{subfigure}[t]{0.45\linewidth}
        \centering
        \includegraphics[width=0.5\linewidth, clip, trim={0cm 0cm 0cm 0cm}]{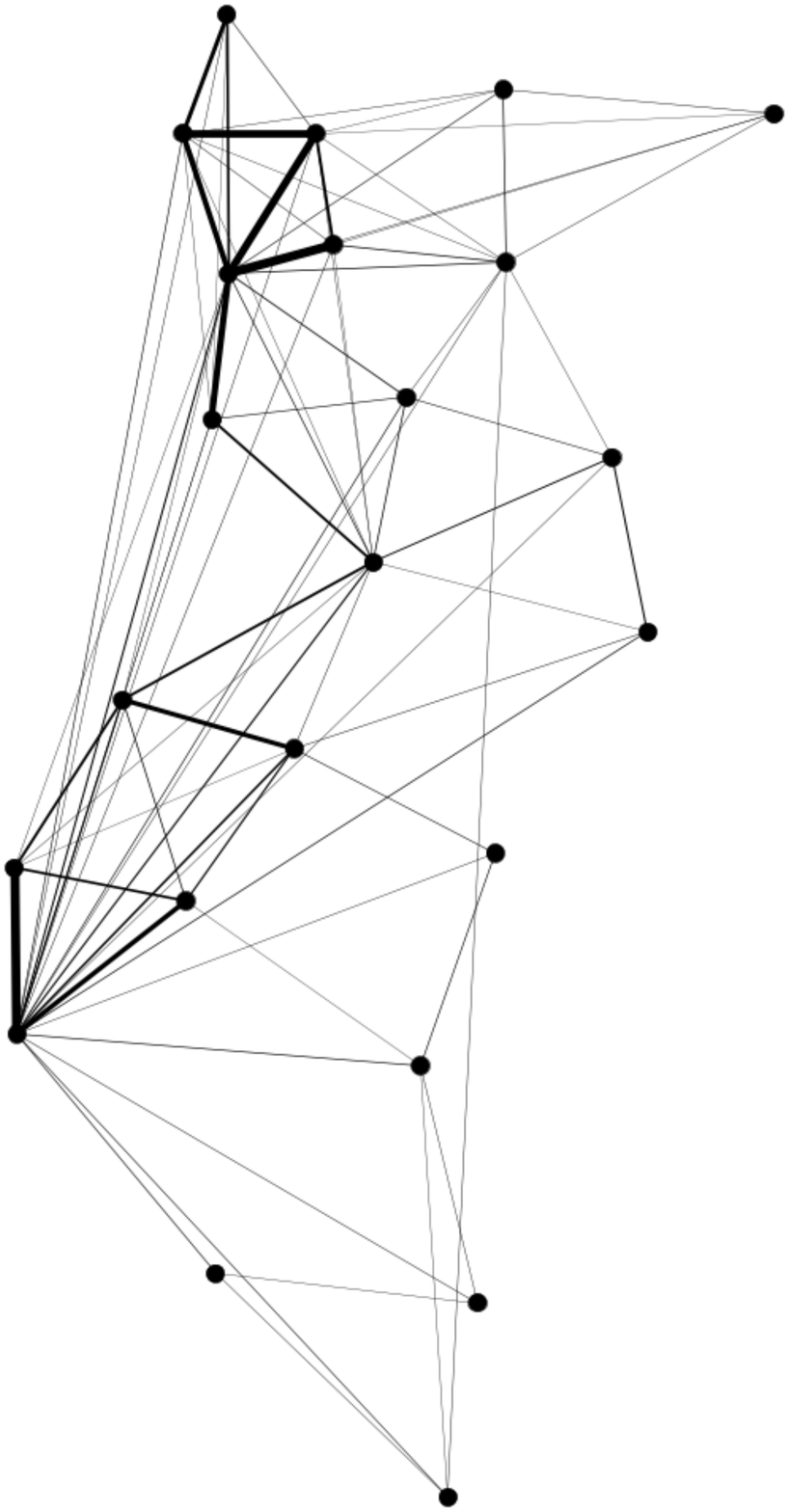}
        \caption{}
        \label{chp7_fig:Map_PTNet}
    \end{subfigure}
    \caption[Geographic representation of Portuguese NUTS 3 and the correspondent weighted network]{(a) Geographic representation of Portuguese NUTS 3, and (b) Topologic representation of the Portuguese NUTS 3 interval-weighted commuters network (IWCN).}
    \label{chp7_fig:Map_PTandNet}
\end{figure}

Figure~\ref{chp7_fig:Map_PTandNet} shows the geographical distribution of NUTS 3 in mainland Portugal (Figure~\ref{chp7_fig:Map_PT}), and the corresponding network of commuting movements between these NUTS 3, weighted by intervals denoting the maximum variability (Figure\ref{chp7_fig:Map_PTNet})\footnote{For the sake of visualization, we chose not to represent the intervals on the network edges, such as it is depicted in Figure~\ref{chp7_fig:directed_to_undirected_edges}d.}. This network has 23 vertices and 80 edges and is therefore considered a small network with low density (considering the intervals midpoints: $\text{graph density}=0.316$, $\text{diameter}=3$, $\text{average degree}=6.96$). For ease of reading, hereinafter we will only refer to Portugal instead of ``mainland Portugal''.

\subsection{Results -- Method 1: Classic Louvain (CL) v.s. Method 2: Hybrid Louvain (HL)}

To assess the outcome of our community detection methodology for interval-weighted networks (IWN) and better understand the effect that these different methods have on the final solution, whether on the number, composition, and value of modularity, in Table~\ref{chp7_tab_1.1:SUM_Diff_Resumo_d1}, we summarize the main results\footnote{It is important to highlight the fact that the numerical values for the different modularities ($Q^{I}$, $Q^{I}_{norm}$ and $Q^{I}_{\max}$) are not comparable since different mathematical procedures are used in each method.}.
The main conclusion is that, despite the equal final number of communities for both methods (three communities), the LA for IWN does not produce the same intermediate (Pass 1) and final (Pass 2) clustering of NUTS 3.
In fact, the communities resulting from the application of the CL method $(Q^{I}_{norm}=0.590)$, tend to roughly represent the division of the country into two major regions, the \textit{northern region} (C\textsubscript{2}: {\footnotesize{AMI, ATA, AMP, AVE, CAV, DOU, RAV, RCO, TES, TTM, VDL}}), and the \textit{southern region} (C\textsubscript{1}: {\footnotesize{ACE, AAL, BAL, ALI, ALG, AML, LTJ, OES, MTJ, RLE}}). The \textit{interior region center} of Portugal (C\textsubscript{3}: {\footnotesize{BBA, BSE}}), forms a residual community on its own. On the contrary, for the HL method, the three NUTS 3 communities $(Q^{I}_{norm}=0.450)$ roughly represent the division of the country into three major regions, the northern region (C\textsubscript{2}: {\small{AMI, AMP, AVE, CAV, RAV, TES, ATA, DOU, TTM}}), the central region (C\textsubscript{3}: {\small{BBA, BSE, RCO, VDL, MTJ, RLE}}), and the southern region (C\textsubscript{1}: {\small{ACE, ALI, ALG, AAL, AML, BAL, LTJ, OES}}).

\begin{table}[H]
\centering
\begin{threeparttable}
\caption{Summary of the outcomes obtained for the Interval-Weighted Commuters Network (IWCN), according to both community detection methods, Classical and Hybrid Louvain.}
\label{chp7_tab_1.1:SUM_Diff_Resumo_d1}
\fontsize{8}{10}\selectfont
\renewcommand{\arraystretch}{1}
\begin{tabular}{M{2.5cm}?M{3cm}??M{3cm}} \thickhline
& \multicolumn{2}{c}{\cellcolor{gray!20} Community Detection Method} \\ \cline{2-3}
	 									& Classic Louvain (CL) 	& Hybrid Louvain (HL) \\ \thickhline
$Q^I$         							& $6371.6$					& $2001.4$  \\ \hline
$Q^I_{max}$   							& $10792.1$         		& $4444.4$  \\ \hline
$Q^I_{norm}$  							& $0.590$           		& $0.450$  	\\ \hline
No. communities   						 & $3$               		& $3$ 		\\ \hline
\multirow{3}{*}{Communities\tnote{a}} & \tiny{ACE, AAL, BAL, ALI, ALG, AML, LTJ, OES, MTJ, RLE} & \tiny{ACE, ALI, ALG, AAL, AML, BAL, LTJ, OES}	 \\ \cline{2-3} 
										& \tiny{AMI, ATA, AMP, AVE, CAV, DOU, RAV, RCO, TES, TTM, VDL} & \tiny{AMI, AMP, AVE, CAV, RAV, TES, ATA, DOU, TTM} \\ \cline{2-3}
                                		& \tiny{BBA,BSE} & \tiny{BBA, BSE, RCO, VDL, MTJ, RLE} \\ \hline
No. Passes                      		& $3$ & $3$ \\ \hhline{===}
\multirow{8}{*}{Pass 1}         		& \scriptsize{$5$ iterations} & \scriptsize{$5$ iterations}\\ \cline{2-3}
								  		& \scriptsize{$Q^I= 3546.6$} & \scriptsize{$Q^I= 2460.9$} \\ \cline{2-3}
                                		& \scriptsize{$6$ communities} & \scriptsize{$5$ communities} \\ \cline{2-3}
								  		& \multirow{6}{*}{} \tiny{ACE, AAL, BAL} & \multirow{5}{*}{} \tiny{ACE, ALI, ALG, AAL, AML, BAL, LTJ, OES} \\ \cline{2-3} 
								  		& \tiny{ALI, ALG} & \tiny{AMI, AMP, AVE, CAV, RAV, TES} \\ \cline{2-3}
								  		& \tiny{AMI, ATA, AMP, AVE, CAV, DOU, RAV, RCO, TES, TTM, VDL} & \tiny{ATA, DOU, TTM} \\ \cline{2-3} 
								  & \tiny{AML, LTJ, OES} & \tiny{BBA, BSE, RCO, VDL}\\ \cline{2-3}
								  & \tiny{BBA, BSE} & \tiny{MTJ, RLE} \\ \cline{2-3}
								  & \tiny{MTJ, RLE} & -- \\ \thickhline
\multirow{3}{*}{Pass 2}         & \scriptsize{$2$ iterations} & \scriptsize{$2$ iterations}  \\ \cline{2-3} 
                                & \scriptsize{$Q^I=6371.6$} & \scriptsize{$Q^I=2001.4$} \\ \cline{2-3} 
                                & \scriptsize{$3$ communities} & \scriptsize{$3$ communities} \\ \thickhline
Pass 3                          & \scriptsize{No change}  & \scriptsize{No change} \\ \thickhline
\end{tabular}%
\begin{tablenotes}
\tiny
   \item [a] {NUTS 3: ACE-Alentejo Central, ALI-Alentejo Litoral, ALG-Algarve, AAL-Alto Alentejo, AMI-Alto Minho, ATA-Alto T\^amega, AML-\'Area Metropolitana de Lisboa, AMP-\'Area Metropolitana do Porto, AVE-Ave, BAL-Baixo Alentejo, BBA-Beira Baixa, BSE-Beiras e Serra da Estrela, CAV-C\'avado, DOU-Douro, LTJ-Lez\'iria do Tejo, MTJ-M\'edio Tejo, OES-Oeste, RAV-Regi\~ao de Aveiro, RCO-Regi\~ao de Coimbra, RLE-Regi\~ao de Leiria, TES-T\^amega e Sousa, TTM-Terras de Tr\'as-os-Montes, VDL-Viseu D\~ao Laf\~oes.}
	\item [\textbullet] Modularity: {$Q^I=\sum_r D\left(o_{rr},e_{rr}\right)$; Normalized modularity: $Q^I_{norm}=\sfrac{Q^I}{Q^I_{max}}$.} 	
	\item [\textbullet] Difference: {$D\big([\underline{x},\overline{x}],[\underline{y},\overline{y}]\big)=\max\big\{|\underline{x}-\underline{y}|,|\overline{x}-\overline{y}|\big\}\times sign\ argmax\big\{|\underline{x}-\underline{y}|,|\overline{x}-\overline{y}|\big\}$.}
	\item [\textbullet] Modularity gain: {$\Delta Q^I= Q^I_{new}-Q^I_{last}$.}
    \end{tablenotes}
  \end{threeparttable}
\end{table}

\bigskip

A useful way to visually distinguish these differences is to employ territorial maps, where NUTS 3 belonging to the same communities are associated with the same shade of gray as depicted below in Figure~\ref{Comparison_CL-HL}. Another useful representation present in Figure~\ref{Comparison_CL-HL} (between the maps) is the \textit{dendrogram}, revealing the hierarchy of the communities (the vertical dashed lines show the current number of communities and their respective ``super-vertices'') showing how Louvain's algorithm clustered the NUTS 3, providing an insight of the pattern of the network.

\newpage

\begin{figure}[H]
    \centering
    \begin{subfigure}[t]{0.35\linewidth}
    \centering
    \resizebox{.70\linewidth}{!}{%
        \begin{tikzpicture}[inner sep=0pt, minimum size=5mm, auto, remember picture]
	\node[node_style,fill=gray!20,font=\fontsize{7}{7}\selectfont,inner sep=0, text width=20mm, align=center] (c2) at (3,2.5)
{\textbf{C\textsubscript{2}}\\ \texttt {AMI, ATA, AMP, AVE, CAV, DOU, RAV, RCO, TES, TTM, VDL}};
	\node[node_style,fill=gray!20,font=\fontsize{7}{7}\selectfont,inner sep=0, text width=18mm, align=center] (c3) at (3,-2.5)
{\textbf{C\textsubscript{1}}\\ \texttt {ACE, AAL, BAL, ALI, ALG,  AML, LTJ, OES, MTJ, RLE}};
	\node[node_style,fill=gray!20,font=\fontsize{7}{7}\selectfont,inner sep=0, text width=12mm, align=center] (c4) at (6,0) {\textbf{C\textsubscript{3}}\\ \texttt {BBA,BSE}};
	\draw[-latex, out=100,in=80,distance=.75cm]  (c2) edge node[above,pos=0.5,font=\fontsize{8}{10}\selectfont] {$[4328, 41994]$} (c2);
	\draw[-latex, out=280,in=260,distance=.75cm]  (c3) edge node[below=0.1cm,pos=0.5,font=\fontsize{8}{10}\selectfont] {$[2562, 24720]$} (c3);
	\draw[-latex, out=10,in=350,distance=.75cm]  (c4) edge node[right=0.1,pos=0.5,font=\fontsize{8}{10}\selectfont] {$[110, 996]$} (c4);
	\draw[edge_style]  (c3) edge node[above,sloped,pos=0.5,font=\fontsize{8}{10}\selectfont] {$[966, 3483]$} (c2);
	\draw[edge_style]  (c3) edge node[below,sloped,pos=0.5,font=\fontsize{8}{10}\selectfont] {$[269, 411]$} (c4);
	\draw[edge_style]  (c4) edge node[above,sloped,pos=0.5,font=\fontsize{8}{10}\selectfont] {$[221, 731]$} (c2);
\end{tikzpicture}
}%
\end{subfigure}
	\begin{subfigure}[t]{0.20\linewidth}
		\centering        
        \includegraphics[width=\linewidth, clip, trim={2.5cm 0cm 2.5cm 0cm}]{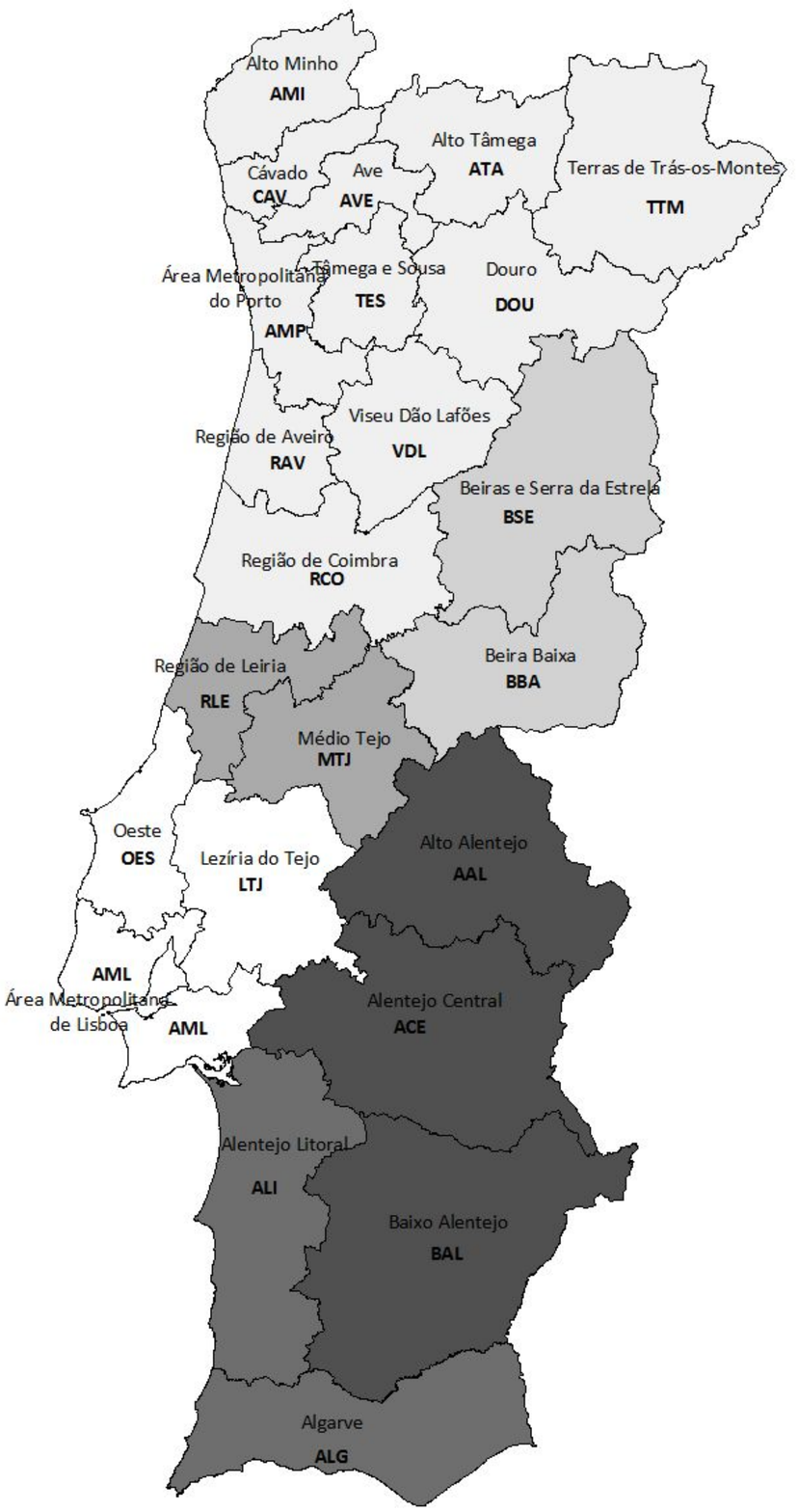}
    \end{subfigure}
%
\begin{subfigure}[t]{0.20\linewidth}
\centering
\resizebox{0.80\linewidth}{!}{%
	\centering
  	\tikzset{
    ncbar angle/.initial=90,
    ncbar/.style={
        to path=(\tikztostart)
        -- ($(\tikztostart)!#1!\pgfkeysvalueof{/tikz/ncbar angle}:(\tikztotarget)$) 
        -- ($(\tikztotarget)!($(\tikztostart)!#1!\pgfkeysvalueof{/tikz/ncbar angle}:(\tikztotarget)$)!\pgfkeysvalueof{/tikz/ncbar angle}:(\tikztostart)$) 
           \tikztonodes
        -- (\tikztotarget) 
        },
    }
\begin{tikzpicture}[font=\tiny,align=right, rotate=270]
	\node (a1) at  (0,    0) {ACE};
	\node (a2) at  (0.40, 0) {AAL};
	\node (a3) at  (0.8,  0) {BAL};
	\node (a4) at  (1.2,  0) {ALI};
	\node (a5) at  (1.6,  0) {ALG};
	\node (a6) at  (2,    0) {AML};
	\node (a7) at  (2.4,  0) {LTJ};
	\node (a8) at  (2.8,  0) {OES};
	\node (a9) at  (3.2,  0) {MTJ};
	\node (a10) at (3.6,  0) {RLE};
	\node (a11) at (4,    0) {AMI};
	\node (a12) at (4.4,  0) {ATA};
	\node (a13) at (4.8,  0) {AMP};
	\node (a14) at (5.2,  0) {AVE};
	\node (a15) at (5.6,  0) {CAV};
	\node (a16) at (6,    0) {DOU};
	\node (a17) at (6.4,  0) {RAV};
	\node (a18) at (6.8,  0) {RCO};
	\node (a19) at (7.2,  0) {TES};
	\node (a20) at (7.6,  0) {TTM};
	\node (a21) at (8,    0) {VAL};
	\node (a22) at (8.4,  0) {BBA};
	\node (a23) at (8.8,  0) {BSE};
	%
	\draw[-] (a1) to [ncbar=3.4] (a2);
	\draw[-] (a2) to [ncbar=3.4] (a3);
	\draw[-] (a4) to [ncbar=3.4] (a5);
	\draw[-] (a6) to [ncbar=3.4] (a7);
	\draw[-] (a7) to [ncbar=3.4] (a8);
	\draw[-] (a9) to [ncbar=3.4] (a10);
	\draw[-] (a11) to [ncbar=3.4] (a12);
	\draw[-] (a12) to [ncbar=3.4] (a13);
	\draw[-] (a13) to [ncbar=3.4] (a14);
	\draw[-] (a14) to [ncbar=3.4] (a15);
	\draw[-] (a15) to [ncbar=3.4] (a16);
	\draw[-] (a16) to [ncbar=3.4] (a17);
	\draw[-] (a17) to [ncbar=3.4] (a18);
	\draw[-] (a18) to [ncbar=3.4] (a19);
	\draw[-] (a19) to [ncbar=3.4] (a20);
	\draw[-] (a20) to [ncbar=3.4] (a21);
	\draw[-] (a22) to [ncbar=3.4] (a23);
	%
	\coordinate (pass11) at (0.4, 1.35);
	\coordinate (pass12) at (1.4, 1.35);
	\draw[-] (pass11) to [ncbar=1.35] (pass12);	
	\coordinate (pass13) at (2.4, 1.35);
	\draw[-] (pass12) to [ncbar=1.35] (pass13);	
	\coordinate (pass14) at (3.4, 1.35);
	\draw[-] (pass13) to [ncbar=1.35] (pass14);		
	%
	\coordinate (pass21) at (1.8, 2.7);
	\coordinate (pass22) at (6, 2.7);
	\draw[-] (pass21) to [ncbar=0.4] (pass22);	
	\coordinate (pass221) at (6.0, 1.35);
	\draw[-] (pass221) to (pass22);
	\coordinate (pass23) at (8.6, 2.7);
	\draw[-] (pass22) to [ncbar=0.646] (pass23);	
	\coordinate (pass231) at (8.6, 1.35);
	\draw[-] (pass231) to (pass23);
	\coordinate (cut1) at (-0.5,  2.1);
	\node [font=\scriptsize] (cut2) at (9.5,2.1) {1\textsuperscript{st} Pass};
	\coordinate (cut3) at (-0.5,  3.6);
	\node [font=\scriptsize] (cut4) at (9.5,3.60) {2\textsuperscript{nd} Pass};
 	\draw [-latex,dashed,rotate=90] (cut1) to (cut2);
 	\draw [-latex,dashed,rotate=90] (cut3) to (cut4);
\end{tikzpicture}
}
       \label{chp7_fig:method_midpoint-dendrograma}
	\end{subfigure}
	\begin{subfigure}[t]{0.2\linewidth}
		\centering        
        \includegraphics[width=\linewidth, clip, trim={2.5cm 0cm 2.5cm 0cm}]{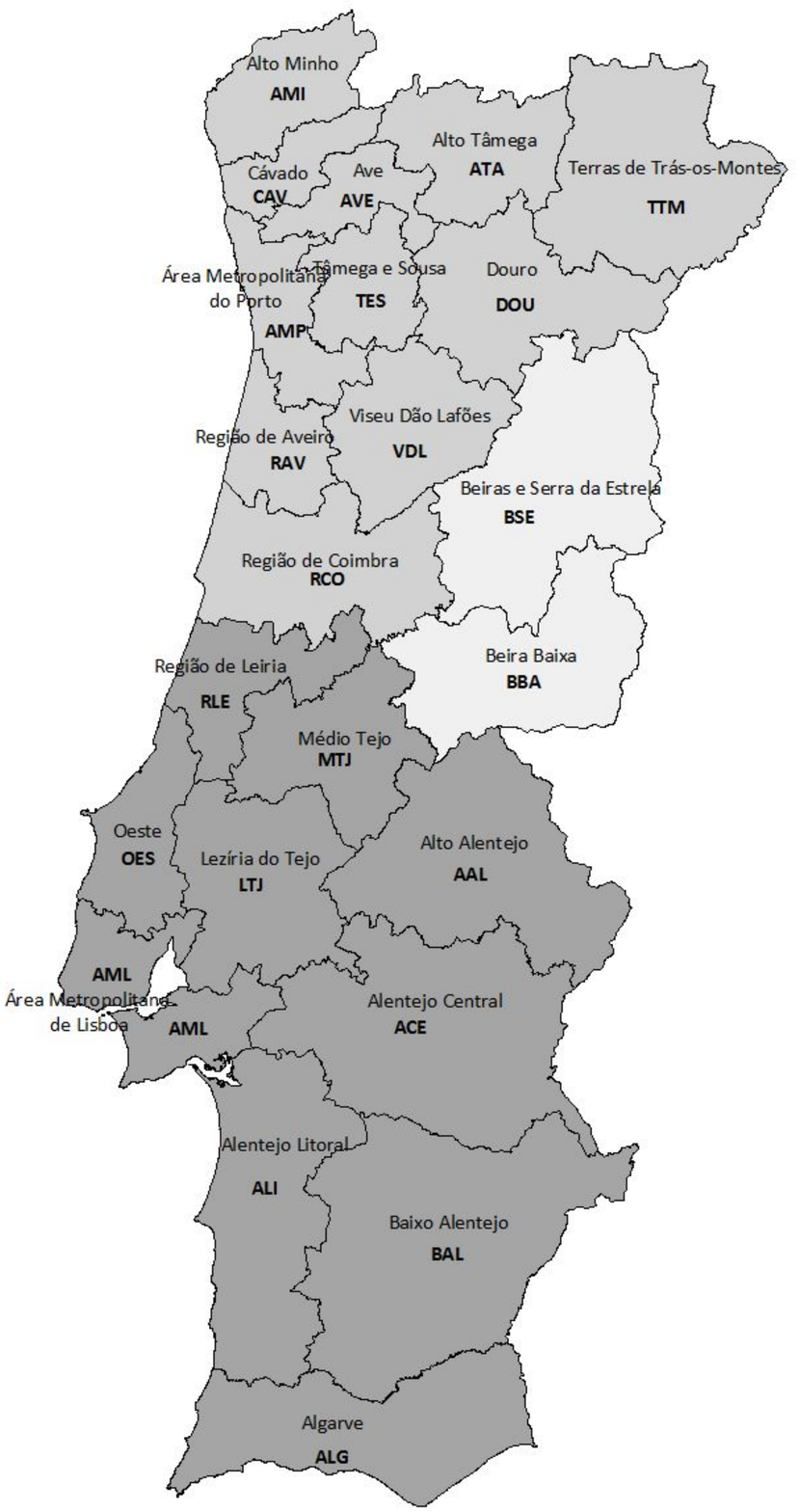}
        \label{chp7_fig:method_midpoint-map2}
    \end{subfigure}
	\caption*{(a) Community structure according to Method 1: ``Classic Louvain (CL)''.}
\vspace{0.5cm}
    
\centering
    \begin{subfigure}[t]{0.35\linewidth}
    \centering
    \resizebox{.75\linewidth}{!}{%
        \begin{tikzpicture}[inner sep=0pt, minimum size=5mm, auto, remember picture]
	\node[node_style,fill=gray!20,font=\fontsize{7}{7}\selectfont,inner sep=0, text width=20mm, align=center] (c2) at (3,2)
{\textbf{C\textsubscript{2}}\\ \texttt {AMI, AMP, AVE, CAV, RAV, TES, ATA, DOU, TTM}};
	\node[node_style,fill=gray!20,font=\fontsize{7}{7}\selectfont,inner sep=0, text width=18mm, align=center] (c3) at (3,-2)
{\textbf{C\textsubscript{1}}\\ \texttt {ACE, ALI, ALG, AAL, AML, BAL, LTJ, OES}};
	\node[node_style,fill=gray!20,font=\fontsize{7}{7}\selectfont,inner sep=0, text width=12mm, align=center] (c4) at (6,0) {\textbf{C\textsubscript{3}}\\ \texttt {BBA, BSE, RCO, VDL, MTJ, RLE}};
	\draw[-latex, out=100,in=80,distance=.75cm]  (c2) edge node[above,pos=0.5,font=\fontsize{9}{10}\selectfont] {$[51, 3398]$} (c2);
	\draw[-latex, out=280,in=260,distance=.75cm]  (c3) edge node[below=0.1cm,pos=0.5,font=\fontsize{9}{10}\selectfont] {$[51, 3340]$} (c3);
	\draw[-latex, out=10,in=350,distance=.75cm]  (c4) edge node[right=0.1,pos=0.5,font=\fontsize{9}{10}\selectfont] {$[51, 1679]$} (c4);
	\draw[edge_style]  (c3) edge node[above,sloped,pos=0.5,font=\fontsize{9}{10}\selectfont] {$[51, 493]$} (c2);
	\draw[edge_style]  (c3) edge node[below,sloped,pos=0.5,font=\fontsize{9}{10}\selectfont] {$[51, 909]$} (c4);
	\draw[edge_style]  (c4) edge node[above,sloped,pos=0.5,font=\fontsize{9}{10}\selectfont] {$[52, 887]$} (c2);
		\end{tikzpicture}
	}%
	\label{chp7_fig:method_hybrid_2.1.1_midpoint_network}
    \end{subfigure}
	\begin{subfigure}[t]{0.20\linewidth}
		\centering        
        \includegraphics[width=\linewidth, clip, trim={2.5cm 0cm 2.5cm 0cm}]{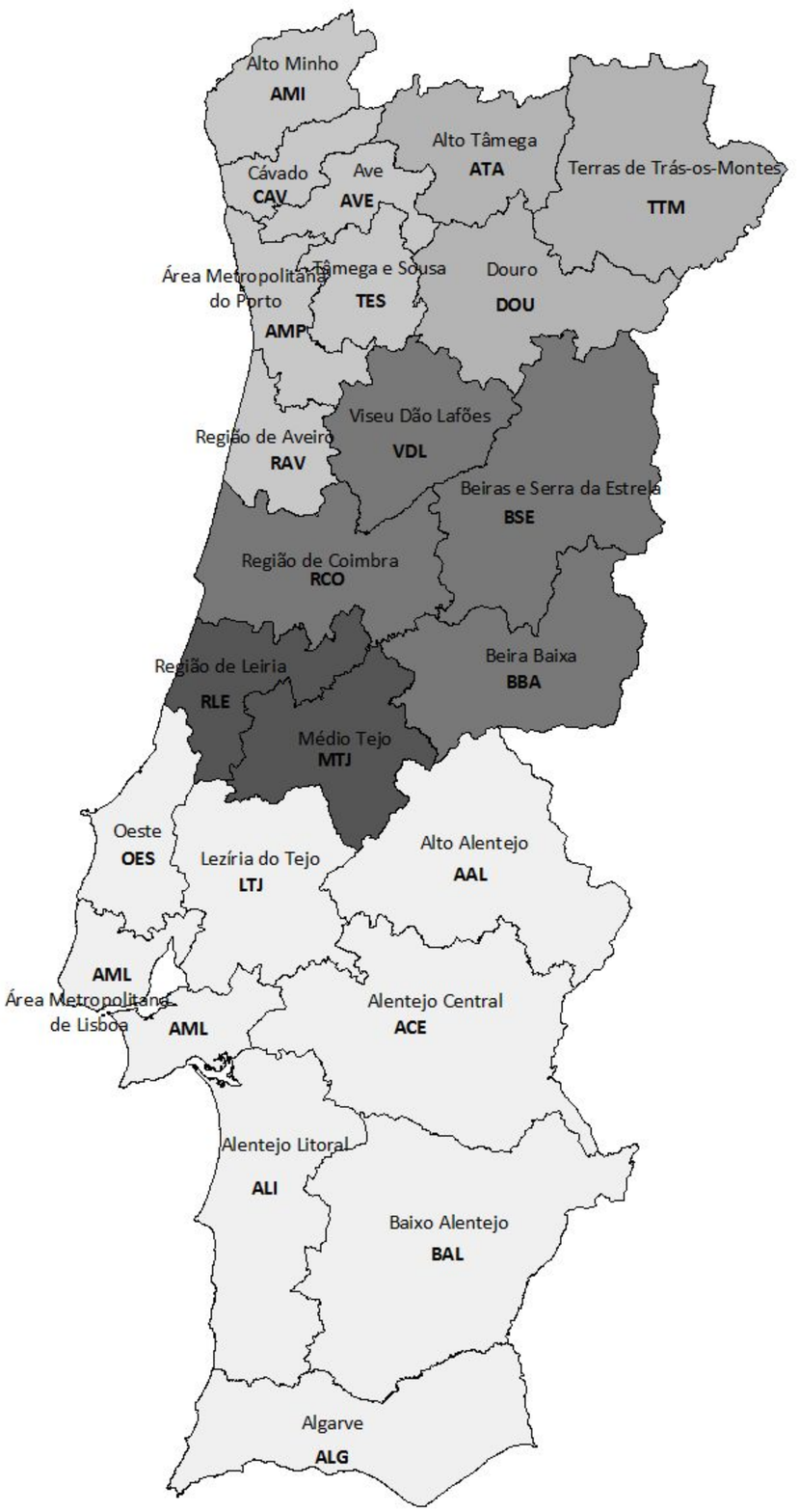}
        \label{chp7_fig:method_hybrid_2.1.1_midpoint_map1}  
    \end{subfigure}
%
\begin{subfigure}[t]{0.20\linewidth}
\centering
\resizebox{0.80\linewidth}{!}{%
	\centering
  	\tikzset{
    ncbar angle/.initial=90,
    ncbar/.style={
        to path=(\tikztostart)
        -- ($(\tikztostart)!#1!\pgfkeysvalueof{/tikz/ncbar angle}:(\tikztotarget)$) 
        -- ($(\tikztotarget)!($(\tikztostart)!#1!\pgfkeysvalueof{/tikz/ncbar angle}:(\tikztotarget)$)!\pgfkeysvalueof{/tikz/ncbar angle}:(\tikztostart)$) 
           \tikztonodes
        -- (\tikztotarget) 
        },
    }
\begin{tikzpicture}[font=\tiny,align=right, rotate=270]
	\node (a1) at  (0,    0) {ACE};
	\node (a2) at  (0.40, 0) {ALI};
	\node (a3) at  (0.8,  0) {ALG};
	\node (a4) at  (1.2,  0) {AAL};
	\node (a5) at  (1.6,  0) {AML};
	\node (a6) at  (2,    0) {BAL};
	\node (a7) at  (2.4,  0) {LTJ};
	\node (a8) at  (2.8,  0) {OES};
	\node (a9) at  (3.2,  0) {AMI};
	\node (a10) at (3.6,  0) {AMP};
	\node (a11) at (4,    0) {AVE};
	\node (a12) at (4.4,  0) {CAV};
	\node (a13) at (4.8,  0) {RAV};
	\node (a14) at (5.2,  0) {TES};
	\node (a15) at (5.6,  0) {ATA};
	\node (a16) at (6,    0) {DOU};
	\node (a17) at (6.4,  0) {TTM};
	\node (a18) at (6.8,  0) {BBA};
	\node (a19) at (7.2,  0) {BSE};
	\node (a20) at (7.6,  0) {RCO};
	\node (a21) at (8,    0) {VDL};
	\node (a22) at (8.4,  0) {MTJ};
	\node (a23) at (8.8,  0) {RLE};
	%
	\coordinate (pass12) at (4.2, 1.35);
	\coordinate (pass13) at (6,   1.35);
	\coordinate (pass14) at (7.4, 1.35);
	\coordinate (pass15) at (8.6, 1.35);
	\coordinate (x) at      (1.4, 1.35);
	\coordinate (passx) at  (1.4, 4.3);
	\coordinate (passx1) at (5, 4.3);
	\coordinate (passx2) at (8,   4.3);
	\coordinate (cut1) at (-0.5,  2.25);
	\node [font=\scriptsize] (cut2) at (9.5,2.25) {1\textsuperscript{st} Pass};
	\coordinate (pass21) at (5,   2.97);
	\coordinate (pass22) at (8,   2.97);
	\coordinate (cut3) at (-0.5,  3.6);
	\node [font=\scriptsize] (cut4) at (9.5,3.60) {2\textsuperscript{nd} Pass};
	\draw[-] (a1) to [ncbar=3.4] (a2);
	\draw[-] (a2) to [ncbar=3.4] (a3);
	\draw[-] (a3) to [ncbar=3.4] (a4);
	\draw[-] (a4) to [ncbar=3.4] (a5);
	\draw[-] (a5) to [ncbar=3.4] (a6);
	\draw[-] (a6) to [ncbar=3.4] (a7);
	\draw[-] (a7) to [ncbar=3.4] (a8);
	\draw[-] (a9) to [ncbar=3.4] (a10);
	\draw[-] (a10) to [ncbar=3.4] (a11);
	\draw[-] (a11) to [ncbar=3.4] (a12);
	\draw[-] (a12) to [ncbar=3.4] (a13);
	\draw[-] (a13) to [ncbar=3.4] (a14);
	\draw[-] (a15) to [ncbar=3.4] (a16);
	\draw[-] (a16) to [ncbar=3.4] (a17);
	\draw[-] (a18) to [ncbar=3.4] (a19);
	\draw[-] (a19) to [ncbar=3.4] (a20);
	\draw[-] (a20) to [ncbar=3.4] (a21);
	\draw[-] (a22) to [ncbar=3.4] (a23);

	\draw[-] (pass12) to [ncbar=0.90] (pass13);
	\draw[-] (pass14) to [ncbar=1.35] (pass15);
	\draw[-] (x) to (passx) to (passx1);
	\draw[-] (pass21) to (passx1) to (passx2) to (pass22);
 	\draw [-latex,dashed,rotate=90] (cut1) to (cut2);
 	\draw [-latex,dashed,rotate=90] (cut3) to (cut4);
\end{tikzpicture}}
       \label{chp7_fig:method_midpoint-dendrograma}
	\end{subfigure}
	\begin{subfigure}[t]{0.2\linewidth}
		\centering        
        \includegraphics[width=\linewidth, clip, trim={2.5cm 0cm 2.5cm 0cm}]{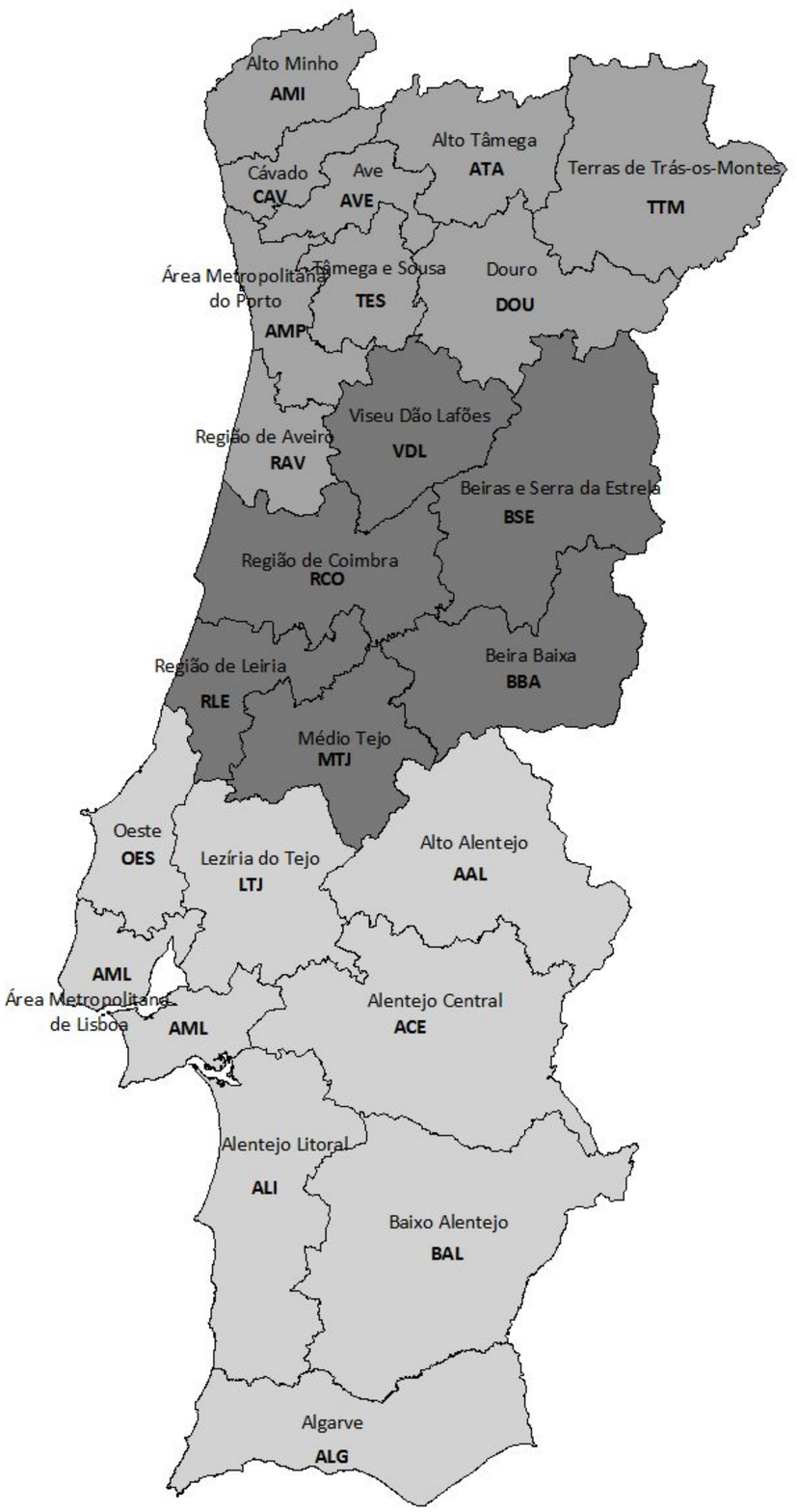}
        \label{chp7_fig:method_midpoint-map2}
    \end{subfigure}
    \caption*{(b) Community structure according to Method 2: ``Hybrid Louvain (HL)''.}
\caption{On the left of the sketch for both methods is represented the final IWN with super-vertices, followed on the right by the geographical representation of communities at level 1 (end of first \textit{pass}) of the LA, the Dendrogram for the Louvain's community detection process and, the geographical representation of communities at level 2 (end of second \textit{pass}) of the LA.}
\label{Comparison_CL-HL}
\end{figure}

\bigskip

Additionally, in Appendix~\ref{appendix_B}, we show the \textit{adjacency matrices} for the IWN obtained from each aggregation method used, which is equivalent to the leftmost pictures (IWN) of Figure~\ref{Comparison_CL-HL}. The intervals account for the maximum variation in daily commuters flows within and between their respective final communities. As expected, the largest variations (between minimum and maximum number of daily commuters) are within their respective communities and the lowest between these communities.

\newpage

\subsection{Discussion}

To evaluate the effect on community detection results of having intervals instead of constants at the edges of an undirected weighted network, and in order to have a basis for comparison, we also apply the Louvain algorithm to the \textit{commuters weighted network} where the weights on the edges correspond to the midpoints of the original intervals. The midpoints of the original intervals correspond to the ``classic'' situation where the weights are constant rather than intervals.
The results are reported in Appendix~\ref{appendix_D}. Considering this method, we may conclude that the final clustering (2\textsuperscript{nd} pass of the algorithm) is very similar to that obtained with the Method 2 ``Hybrid Louvain (HL)''. The only change in communities composition occurs in the transfer of ``M\'{e}dio Tejo'' ({\small{MTJ}}) and ``Regi\~{a}o de Leiria ({\small{RLE}}) from community 3 to community 1. This is to be expected, since both methods use the midpoints to evaluate modularity gains and decide which vertices should be merged.\\
On the other hand, Method 1 (CL), which considers information in the form of intervals, thus better capturing the variability present in the raw data, tends to divide the national territory according to commuters mobility in the context of the country's territorial density. In this way, it forms broader territorial communities that accompany the country's population density, namely, the entire North region, where the population density is higher, the Center/South regions and clearly isolating the Interior Center region (``Beiras'') with less population density.\\
However, the final adjacency matrix within and between communities in the form of intervals obtained by both Methods (CL and HL), is richer than the one produced by the Louvain method for weighted networks, since it provides information about the variability of the commuters movements within and between communities.\\

\section{Concluding remarks}
\label{conclusion}

In this paper, we present a new methodology to detect the community structure that emerges from an interval-weighted network (IWN), based on two different methods, which we name ``Method 1: Classic Louvain (CL)'' and ``Method 2: Hybrid Louvain (HL)''. In the former, both the \textit{optimization} on 1\textsuperscript{st} phase and the \textit{aggregation} of the network on 2\textsuperscript{nd} phase, are calculated by \textit{summing} the intervals for the formed communities. In the latter, the \textit{optimization} on 1\textsuperscript{st} phase is performed by \textit{using the midpoints} of the intervals, and in \textit{2\textsuperscript{nd} phase, the aggregation} of the network is done by selecting the minimum value and the maximum values of the intervals for the formed communities.\\
Interval-weighted networks (IWN), are characterized by having interval variations (ranges) on the edges, allowing taking into account the variability observed in the original data, thereby minimizing the loss of information. We have shown that an IWN can be represented in the form of an interval-weighted contingency table for the observed and expected intervals. Subsequently, we propose the generalization of modularity $(Q^I)$ and modularity gain $(\Delta Q^I)$ to the case of an IWN. These generalizations are not straightforward, essentially because of the limitations of interval computations. To contour these drawbacks we propose a difference based on the Hausdorff distance but does take into account the sign of the highest value to evaluate de difference between two intervals.\\
We apply our methodology in a real-world commuter network to detect the community structure of movements of the daily commuters in mainland Portugal between the twenty three NUTS 3 Regions. The main conclusion is that the community detection methodology is able to profile homogeneous and contiguous clusters of regions, taking into account the variability of the edges weights. Another important note to highlight, is that these results put in evidence that, according to the method used, despite the same number of final communities, the hierarchy of the communities is different in both Passes of the Louvain algorithm. Apparently, the ``Method 1: Classic Louvain (CL)'' tends to form broader communities than the``Method 2: Hybrid Louvain (HL)''. It is also important to highlight that for the "Method 2 (HL)", since it uses intervals' midpoint in the calculations of the LA optimization phase, the speed of computation is higher, thus allowing for its use in large networks.

The present study may be useful in practical applications based on community detection considering the strength variation and topology of the commuting patterns, specially in territorial studies.

This paper is one of the first attempts in relating interval arithmetic and network analysis. Our findings suggest that further analysis should be developed. First, these methods need to be validated with other territorial data, for example, more desegregated information like, for example, municipalities instead of NUTS 3~\citep{DeMontis:2013ho}. Second, extending our methodology considering the direction between the edges of the interval-weighted network (direct interval-weighted network), or even consider applying algorithms allowing overlapping communities~\citep{Palla:2005cj}.

\paragraph{Acknowledgements:}
This work was financed by the Portuguese funding agency,\linebreak FCT - Funda\c{c}\~ao para a Ci\^encia e a Tecnologia, within project UIDB/50014/2020.
This research has also received funding from the European Union's Horizon 2020 research and innovation program ''FIN-TECH: A Financial supervision and Technology compliance training programme'' under the grant agreement No 825215 (Topic: ICT-35-2018, Type of action: CSA).

\renewcommand{\refname}{\normalsize References}
\bibliographystyle{chicago}
\bibliography{references}


\newpage

\appendix
\section*{Appendix A: \textit{R} output for Method 1. Classic Louvain (CL) -- Intervals Sum}
\label{appendix_C}
\vspace{-0.5cm}
\tiny{
\begin{lstlisting}[basicstyle=\small\linespread{0.75}\listingsfont, frame=single,
caption={\textit{R} output for the CL: Method 1. Intervals Sum $Q^I\ /\ D\ /\ \Delta Q^I$},label=chp5_Appdix_Classic-1]
Initial Interval-Weighted Network:
     v1    v2    v3    v4 
v1 [0,0] [1,3] [1,1] [0,0] 
v2 [1,3] [0,0] [1,1] [0,0] 
v3 [1,1] [1,1] [0,0] [2,4] 
v4 [0,0] [0,0] [2,4] [0,0] 

* Initial Modularity=-7.000
* Begin Pass number 1
	Try v1 -> v2         | gain=+4.095 (+)
	Try v1 -> v3         | gain=-0.810 (-)
	Move v1 -> v2        
	Try v2 -> v1,v2      | gain=+4.095 (+)
	Try v2 -> v3         | gain=-0.810 (-)
	Keep vertex v2 at community v1,v2
	Try v3 -> v1,v2      | gain=-0.679 (-)
	Try v3 -> v4         | gain=+5.762 (+)
	Move v3 -> v4        
	Try v4 -> v3,v4      | gain=+5.762 (+)
	Keep vertex v4 at community v3,v4
Iteration 1 Modularity=2.857
	Try v1 -> v1,v2      | gain=+4.095 (+)
	Try v1 -> v3,v4      | gain=-2.345 (-)
	Keep vertex v1 at community v1,v2
	Try v2 -> v1,v2      | gain=+4.095 (+)
	Try v2 -> v3,v4      | gain=-2.345 (-)
	Keep vertex v2 at community v1,v2
	Try v3 -> v1,v2      | gain=-0.679 (-)
	Try v3 -> v3,v4      | gain=+5.762 (+)
	Keep vertex v3 at community v3,v4
	Try v4 -> v3,v4      | gain=+5.762 (+)
	Keep vertex v4 at community v3,v4
Iteration 2 Modularity=2.857

New network: ---------------
        v1,v2  v3,v4 
v1,v2   [2,6]  [2,2] 
v3,v4   [2,2]  [4,8] 
* End Pass number 1 Modularity=2.857 Communities=v1,v2 / v3,v4
---------------------------
* Begin Pass number 2
	Try v1,v2 -> v1,v2      | gain=+0.000 (0)
	Try v1,v2 -> v3,v4      | gain=-2.857 (-)
	Keep vertex v1,v2 at community v1,v2
	Try v3,v4 -> v1,v2      | gain=-2.857 (-)
	Try v3,v4 -> v3,v4      | gain=+0.000 (0)
	Keep vertex v3,v4 at community v3,v4
Iteration 1 Modularity=2.857
* End Pass number 2 -- no change

* Final communities: v1,v2 / v3,v4 (n=2)
* Before Normalized: 2.857
* Normalized modularity: 0.455 (Qmax=6.285714)
---------------------------
Final Interval-weighted network:

        v1,v2  v3,v4 
v1,v2   [2,6]  [2,2] 
v3,v4   [2,2]  [4,8] 
\end{lstlisting}
}
\normalsize
\newpage

\section*{Appendix B: \textit{R} output for Method 2. Hybrid Louvain (HL) -- Intervals Midpoint}
\label{appendix_A}
\vspace{-0.5cm}
\tiny{
\begin{lstlisting}[basicstyle=\small\linespread{0.75}\listingsfont, frame=single,
caption={\textit{R} output for the LA: Method 2. Intervals Midpoint $Q^I\ /\ D\ /\ \Delta Q^I$},label=chp5_Appdix_Hybrid-2.1]
Initial Interval-Weighted Network:
     v1    v2    v3    v4 
v1 [0,0] [1,3] [1,1] [0,0] 
v2 [1,3] [0,0] [1,1] [0,0] 
v3 [1,1] [1,1] [0,0] [2,4] 
v4 [0,0] [0,0] [2,4] [0,0] 

* Initial Modularity=-3.714
* Begin Pass number 1
	Try v1 -> v2         | gain=+2.714 (+)
	Try v1 -> v3         | gain=-0.143 (-)
	Move v1 -> v2        
	Try v2 -> v1,v2      | gain=+2.714 (+)
	Try v2 -> v3         | gain=-0.143 (-)
	Keep vertex v2 at community v1,v2
	Try v3 -> v1,v2      | gain=-0.286 (-)
	Try v3 -> v4         | gain=+3.857 (+)
	Move v3 -> v4        
	Try v4 -> v3,v4      | gain=+3.857 (+)
	Keep vertex v4 at community v3,v4
Iteration 1 Modularity=2.857
	Try v1 -> v1,v2      | gain=+2.714 (+)
	Try v1 -> v3,v4      | gain=-1.429 (-)
	Keep vertex v1 at community v1,v2
	Try v2 -> v1,v2      | gain=+2.714 (+)
	Try v2 -> v3,v4      | gain=-1.429 (-)
	Keep vertex v2 at community v1,v2
	Try v3 -> v1,v2      | gain=-0.286 (-)
	Try v3 -> v3,v4      | gain=+3.857 (+)
	Keep vertex v3 at community v3,v4
	Try v4 -> v3,v4      | gain=+3.857 (+)
	Keep vertex v4 at community v3,v4
Iteration 2 Modularity=2.857

New network: ---------------
       v1,v2  v3,v4 
v1,v2  [1,3]  [1,1] 
v3,v4  [1,1]  [2,4]  
* End Pass number 1 Modularity=1.429 Communities=v1,v2 / v3,v4
---------------------------
* Begin Pass number 2
	Try v1,v2 -> v1,v2      | gain=+0.000 (0)
	Try v1,v2 -> v3,v4      | gain=-1.429 (-)
	Keep vertex v1,v2 at community v1,v2
	Try v3,v4 -> v1,v2      | gain=-1.429 (-)
	Try v3,v4 -> v3,v4      | gain=+0.000 (0)
	Keep vertex v3,v4 at community v3,v4
Iteration 1 Modularity=1.429
* End Pass number 2 -- no change

* Final communities: v1,v2 / v3,v4 (n=2)
* Hybrid - Before Normalized: 1.429
* Normalized modularity: 0.417 (Qmax=3.428571)
---------------------------
Final Interval-weighted network:

       v1,v2  v3,v4 
v1,v2  [1,3]  [1,1] 
v3,v4  [1,1]  [2,4]  
\end{lstlisting}
}
\normalsize
\newpage


\section*{Appendix C: Adjacency matrices for the interval-weighted network (IWN) obtained from the aggregation method used.}
\label{appendix_B}


\bigskip

\noindent
Method 1. Classic Louvain (CL)

\bigskip

\begin{table}[H]
\centering
\begin{threeparttable}
\caption{Interval-weighted adjacency matrix for the three communities $(C_1,C_2\ \text{and}\ C_3)$\tnote{a} -- Method 1. Classic Louvain (CL).}
\label{tab1}
\fontsize{8}{10}\selectfont
\renewcommand{\arraystretch}{2}
\begin{tabular}{M{2.5cm}|M{2.5cm}|M{2.5cm}|M{2cm}}
    & \makecell{\cellcolor{gray!60}\scriptsize{\textbf{C\textsubscript{1}}}} & \makecell{\cellcolor{gray!30}\scriptsize{\textbf{C\textsubscript{2}}}} & \makecell{\cellcolor{gray!15}\scriptsize{\textbf{C\textsubscript{3}}}}\\
    & \tiny{ACE, AAL, BAL, ALI, ALG, AML, LTJ, OES, MTJ, RLE}   & \tiny{AMI, ATA, AMP, AVE, CAV, DOU, RAV, RCO, TES, TTM, VDL} & \tiny{BBA, BSE} \\\hline
   \tiny{ACE, AAL, BAL, ALI, ALG, AML, LTJ, OES, MTJ, RLE}  & \cellcolor{gray!60}$[2562,24720]$ & $[966, 3483]$ & $[269, 411]$ \\\hline
  \tiny{AMI, ATA, AMP, AVE, CAV, DOU, RAV, RCO, TES, TTM, VDL} & $[966, 3483]$ & \cellcolor{gray!30}$[4328,41994]$ & $[221,731]$ \\\hline
  \tiny{BBA, BSE} & $[269, 411]$ & $[221,731]$ & \cellcolor{gray!15}$[110,996]$ \\
\end{tabular}
\begin{tablenotes}
      \tiny
      \item [a] {NUTS 3: ACE-Alentejo Central, ALI-Alentejo Litoral, ALG-Algarve, AAL-Alto Alentejo, AMI-Alto Minho, ATA-Alto T\^amega, AML-\'Area Metropolitana de Lisboa, AMP-\'Area Metropolitana do Porto, AVE-Ave, BAL-Baixo Alentejo, BBA-Beira Baixa, BSE-Beiras e Serra da Estrela, CAV-C\'avado, DOU-Douro, LTJ-Lez\'iria do Tejo, MTJ-M\'edio Tejo, OES-Oeste, RAV-Regi\~ao de Aveiro, RCO-Regi\~ao de Coimbra, RLE-Regi\~ao de Leiria, TES-T\^amega e Sousa, TTM-Terras de Tr\'as-os-Montes, VDL-Viseu D\~ao Laf\~oes.}
    \end{tablenotes}
  \end{threeparttable}
\end{table}

\bigskip

\noindent
Method 2. Hybrid Louvain (HL)

\begin{table}[H]
\centering
\begin{threeparttable}
\caption{Interval-weighted adjacency matrix for the three communities $(C_1,C_2\ \text{and}\ C_3)$\tnote{a} -- Method 2. Hybrid Louvain (HL).}
\label{chp7_tab:method_hybrid_2.1.1_midpoint_adjc-matrix}
\fontsize{8}{10}\selectfont
\renewcommand{\arraystretch}{2}
\begin{tabular}{M{2.5cm}|M{2.5cm}|M{2.5cm}|M{2cm}}
    & \makecell{\cellcolor{gray!60}\scriptsize{\textbf{C\textsubscript{1}}}} & \makecell{\cellcolor{gray!30}\scriptsize{\textbf{C\textsubscript{2}}}} & \makecell{\cellcolor{gray!15}\scriptsize{\textbf{C\textsubscript{3}}}}\\
  & \tiny{ACE, ALI, ALG, AAL, AML, BAL, LTJ, OES}   & \tiny{AMI, AMP, AVE, CAV, RAV, TES, ATA, DOU, TTM} & \tiny{BBA, BSE, RCO, VDL, MTJ, RLE} \\\hline
\tiny{ACE, ALI, ALG, AAL, AML, BAL, LTJ, OES}  & \cellcolor{gray!60}$[51, 3340]$ & $[51, 493]$ & $[51, 909]$ \\\hline
\tiny{AMI, AMP, AVE, CAV, RAV, TES, ATA, DOU, TTM} & $[51, 493]$ & \cellcolor{gray!30}$[51, 3398]$ & $[52, 887]$ \\\hline
\tiny{BBA, BSE, RCO, VDL, MTJ, RLE} & $[51, 909]$ & $[52, 887]$ & \cellcolor{gray!15}$[51, 1679]$ \\
\end{tabular}
\begin{tablenotes}
      \tiny
      \item [a] {NUTS 3: ACE-Alentejo Central, ALI-Alentejo Litoral, ALG-Algarve, AAL-Alto Alentejo, AMI-Alto Minho, ATA-Alto T\^amega, AML-\'Area Metropolitana de Lisboa, AMP-\'Area Metropolitana do Porto, AVE-Ave, BAL-Baixo Alentejo, BBA-Beira Baixa, BSE-Beiras e Serra da Estrela, CAV-C\'avado, DOU-Douro, LTJ-Lez\'iria do Tejo, MTJ-M\'edio Tejo, OES-Oeste, RAV-Regi\~ao de Aveiro, RCO-Regi\~ao de Coimbra, RLE-Regi\~ao de Leiria, TES-T\^amega e Sousa, TTM-Terras de Tr\'as-os-Montes, VDL-Viseu D\~ao Laf\~oes.}
    \end{tablenotes}
  \end{threeparttable}
\end{table}

\section*{Appendix D: Community structure according to Louvain's Method -- Degenerate Intervals of midpoints}
\label{appendix_D}

\begin{table}[H]
\centering
\begin{threeparttable}
\caption{Summary of the outcomes obtained for the weighted Commuters Network (intervals midpoint), according to Louvain's algorithm.}
\label{chp7_Appendix_tab_Classical-Louvain_Midpoint}
\fontsize{8}{10}\selectfont
\renewcommand{\arraystretch}{0.8}
\begin{tabular}{M{3cm}|M{6cm}}\thickhline
& \multicolumn{1}{c}{\cellcolor{gray!20} Louvain's Method for weighted networks} \\ \thickhline
$Q^N$         & $17255.2$ \\ \hline
$Q^N_{max}$                     & $24961.2$ \\ \hline
$Q^N_{norm}$  & $0.691$ \\ \hline
No. comm.                       & $3$   \\ \hline
\multirow{3}{*}{Communities\tnote{a}} & \tiny{ACE, ALI, ALG, AAL, AML, BAL, LTJ, OES, MTJ, RLE} \\ \cline{2-2} 
										& \tiny{AMI, AMP, AVE, CAV, RAV, TES, ATA, DOU, TTM} \\ \cline{2-2}
                                & \tiny{BBA, BSE, RCO, VDL} \\ \hline
No. Passes                      & $3$  \\ \hhline{==}
\multirow{8}{*}{Pass 1}         & \scriptsize{$5$ iterations} \\ \cline{2-2}
								  & \scriptsize{$Q^N= 15982.0$} \\ \cline{2-2}
                                & \scriptsize{$5$ communities} \\ \cline{2-2}
								  & \multirow{5}{*}{} \tiny{ACE, ALI, ALG, AAL, AML, BAL, LTJ, OES} \\ \cline{2-2} 
								  & \tiny{AMI, AMP, AVE, CAV, RAV, TES} \\ \cline{2-2}
								  & \tiny{ATA, DOU, TTM} \\ \cline{2-2} 
								  & \tiny{BBA, BSE, RCO, VDL} \\ \cline{2-2}
								  & \tiny{MTJ, RLE} \\ \thickhline
\multirow{3}{*}{Pass 2}         & \scriptsize{$2$ iterations}  \\ \cline{2-2} 
                                & \scriptsize{$Q^N=17255.2$} \\ \cline{2-2} 
                                & \scriptsize{$3$ communities}  \\ \thickhline
Pass 3                          & \scriptsize{No change}  \\ \thickhline
\end{tabular}%
\begin{tablenotes}
      \tiny
      \item [a] {NUTS 3: ACE-Alentejo Central, ALI-Alentejo Litoral, ALG-Algarve, AAL-Alto Alentejo, AMI-Alto Minho, ATA-Alto T\^amega, AML-\'Area Metropolitana de Lisboa, AMP-\'Area Metropolitana do Porto, AVE-Ave, BAL-Baixo Alentejo, BBA-Beira Baixa, BSE-Beiras e Serra da Estrela, CAV-C\'avado, DOU-Douro, LTJ-Lez\'iria do Tejo, MTJ-M\'edio Tejo, OES-Oeste, RAV-Regi\~ao de Aveiro, RCO-Regi\~ao de Coimbra, RLE-Regi\~ao de Leiria, TES-T\^amega e Sousa, TTM-Terras de Tr\'as-os-Montes, VDL-Viseu D\~ao Laf\~oes.}
      \item [.] {Modularity: $Q^N=\sum_{C\in \mathcal{C}} \sum_{i,j\in C} \left(o_{ij}-e_{ij}\right)$.}
\end{tablenotes}
\end{threeparttable}
\end{table}

\begin{figure}[H]
    \centering
    \begin{subfigure}[t]{0.35\linewidth}
    \centering
    \resizebox{.70\linewidth}{!}{%
        \begin{tikzpicture}[inner sep=0pt, minimum size=5mm, auto, remember picture]
	\node[node_style,fill=gray!20,font=\fontsize{7}{7}\selectfont,inner sep=0, text width=20mm, align=center] (c2) at (3,2)
{\textbf{C\textsubscript{2}}\\ \texttt {AMI, AMP, AVE, CAV, RAV, TES, ATA, DOU, TTM}};
	\node[node_style,fill=gray!20,font=\fontsize{7}{7}\selectfont,inner sep=0, text width=18mm, align=center] (c3) at (3,-2)
{\textbf{C\textsubscript{1}}\\ \texttt {ACE, ALI, ALG, AAL, AML, BAL, LTJ, OES, MTJ, RLE}};
	\node[node_style,fill=gray!20,font=\fontsize{7}{7}\selectfont,inner sep=0, text width=12mm, align=center] (c4) at (6,0) {\textbf{C\textsubscript{3}}\\ \texttt {BBA, BSE, RCO, VDL}};
	\draw[-latex, out=100,in=80,distance=.75cm]  (c2) edge node[above,pos=0.5,font=\fontsize{7}{10}\selectfont] {$20350$} (c2);
	\draw[-latex, out=280,in=260,distance=.75cm]  (c3) edge node[below=0.1cm,pos=0.5,font=\fontsize{7}{10}\selectfont] {$13641$} (c3);
	\draw[-latex, out=10,in=350,distance=.75cm]  (c4) edge node[right=0.1,pos=0.5,font=\fontsize{7}{10}\selectfont] {$1739$} (c4);
	\draw[edge_style]  (c3) edge node[above,sloped,pos=0.5,font=\fontsize{7}{10}\selectfont] {$1172$} (c2);
	\draw[edge_style]  (c3) edge node[below,sloped,pos=0.5,font=\fontsize{7}{10}\selectfont] {$1392.5$} (c4);
	\draw[edge_style]  (c4) edge node[above,sloped,pos=0.5,font=\fontsize{7}{10}\selectfont] {$1288.5$} (c2);
\end{tikzpicture}
}
\label{chp7_Appendix__method_midpoint_network}
    \end{subfigure}
	\begin{subfigure}[t]{0.20\linewidth}
		\centering        
        \includegraphics[width=\linewidth, clip, trim={2.5cm 0cm 2.5cm 0cm}]{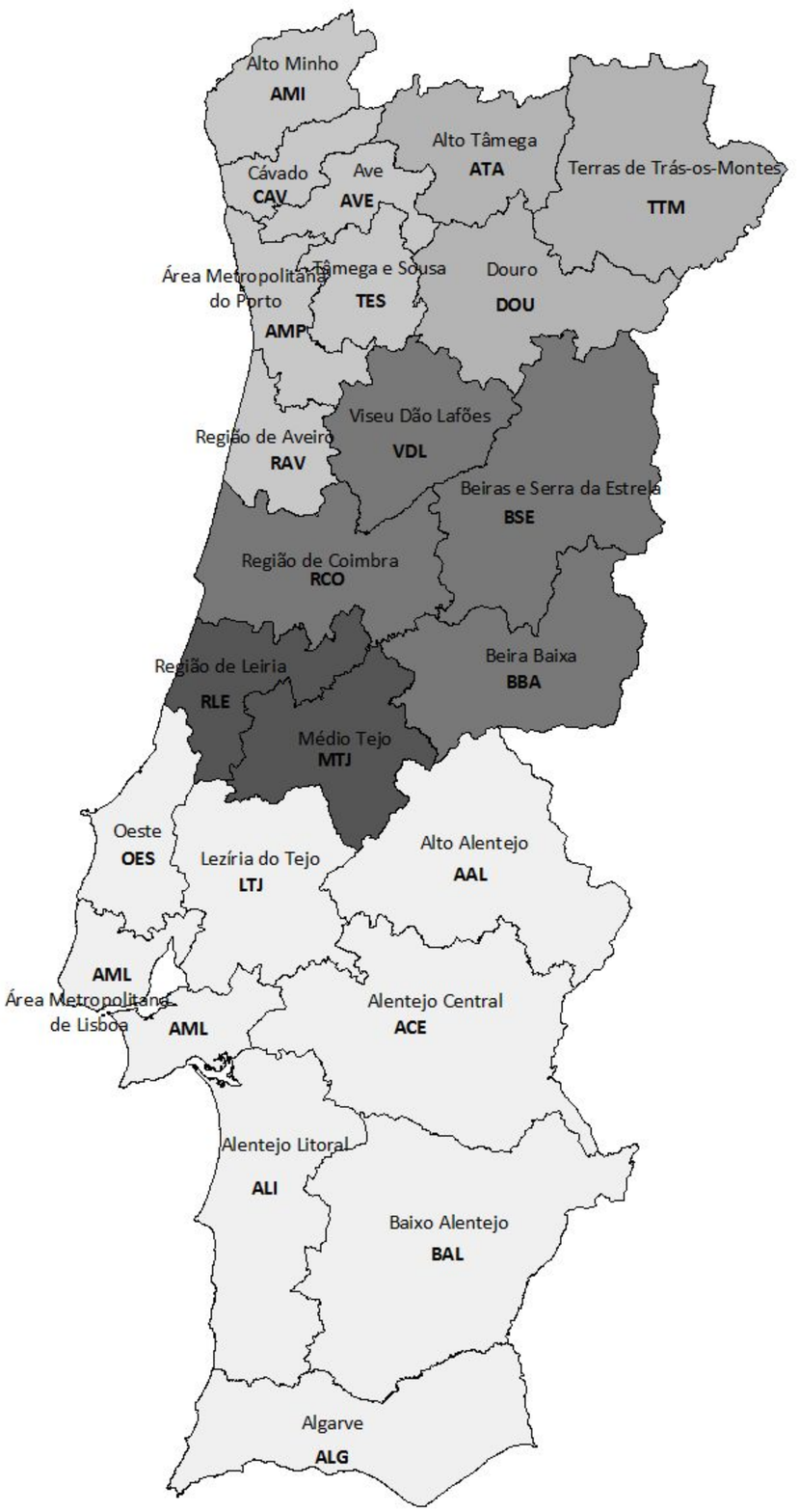}
        \label{chp7_Appendix_fig:method_midpoint-map1}
    \end{subfigure}
%
%
\begin{subfigure}[t]{0.2\linewidth}
\centering
\resizebox{0.80\linewidth}{!}{%
	\centering
  	\tikzset{
    ncbar angle/.initial=90,
    ncbar/.style={
        to path=(\tikztostart)
        -- ($(\tikztostart)!#1!\pgfkeysvalueof{/tikz/ncbar angle}:(\tikztotarget)$) 
        -- ($(\tikztotarget)!($(\tikztostart)!#1!\pgfkeysvalueof{/tikz/ncbar angle}:(\tikztotarget)$)!\pgfkeysvalueof{/tikz/ncbar angle}:(\tikztostart)$) 
           \tikztonodes
        -- (\tikztotarget) 
        },
    }
\begin{tikzpicture}[font=\tiny,align=right, rotate=270]
	\node (a1) at  (0,    0) {ACE};
	\node (a2) at  (0.40, 0) {ALI};
	\node (a3) at  (0.8,  0) {ALG};
	\node (a4) at  (1.2,  0) {AAL};
	\node (a5) at  (1.6,  0) {AML};
	\node (a6) at  (2,    0) {BAL};
	\node (a7) at  (2.4,  0) {LTJ};
	\node (a8) at  (2.8,  0) {OES};
	\node (a9) at  (3.2,  0) {MTJ};
	\node (a10) at (3.6,  0) {RLE};
	\node (a11) at (4,    0) {AMI};
	\node (a12) at (4.4,  0) {AMP};
	\node (a13) at (4.8,  0) {AVE};
	\node (a14) at (5.2,  0) {CAV};
	\node (a15) at (5.6,  0) {RAV};
	\node (a16) at (6,    0) {TES};
	\node (a17) at (6.4,  0) {ATA};
	\node (a18) at (6.8,  0) {DOU};
	\node (a19) at (7.2,  0) {TTM};
	\node (a20) at (7.6,  0) {BBA};
	\node (a21) at (8,    0) {BSE};
	\node (a22) at (8.4,  0) {RCO};
	\node (a23) at (8.8,  0) {VDL};
	\draw[-] (a1) to [ncbar=1.25cm] (a2);
	\draw[-] (a2) to [ncbar=1.25cm] (a3);
	\draw[-] (a3) to [ncbar=1.25cm] (a4);
	\draw[-] (a4) to [ncbar=1.25cm] (a5);
	\draw[-] (a5) to [ncbar=1.25cm] (a6);
	\draw[-] (a6) to [ncbar=1.25cm] (a7);
	\draw[-] (a7) to [ncbar=1.25cm] (a8);
	\draw[-] (a9) to [ncbar=1.25cm] (a10);
	\draw[-] (a11) to [ncbar=1.25cm] (a12);
	\draw[-] (a12) to [ncbar=1.25cm] (a13);
	\draw[-] (a13) to [ncbar=1.25cm] (a14);
	\draw[-] (a14) to [ncbar=1.25cm] (a15);
	\draw[-] (a15) to [ncbar=1.25cm] (a16);
	\draw[-] (a17) to [ncbar=1.25cm] (a18);
	\draw[-] (a18) to [ncbar=1.25cm] (a19);
	\draw[-] (a20) to [ncbar=1.25cm] (a21);
	\draw[-] (a21) to [ncbar=1.25cm] (a22);
	\draw[-] (a22) to [ncbar=1.25cm] (a23);	
	\coordinate (x1) at (1.4, 1.25);
	\coordinate (x2) at (3.4, 1.25);	
	\draw[-] (x1) to [ncbar=1.25cm] (x2);	
	\coordinate (x3) at (5, 1.25);
	\coordinate (x4) at (6.8, 1.25);	
	\draw[-] (x3) to [ncbar=1.25cm] (x4);	
	\coordinate (x5) at (5.9, 2.5);
	\coordinate (x6) at (8.2, 2.5);	
	\draw[-] (x5) to [ncbar=1.25cm] (x6);	
	\coordinate (x7) at (8.2, 1.25);
	\coordinate (x8) at (8.2, 2.5);			
	\draw[-] (x7) to (x8);	
	\coordinate (x9) at (2.4, 2.5);
	\coordinate (x10) at (5.9, 2.5);
	\draw[-] (x9) to [ncbar=1.25cm] (x10);		
	\coordinate (cut1) at (-0.5,  1.8);
	\node [font=\scriptsize] (cut2) at (9.5,1.8) {1\textsuperscript{st} Pass};
	\coordinate (cut3) at (-0.5,  3.25);
	\node [font=\scriptsize] (cut4) at (9.5,3.25) {2\textsuperscript{nd} Pass};
 	\draw [-latex,dashed,rotate=90] (cut1) to (cut2);
 	\draw [-latex,dashed,rotate=90] (cut3) to (cut4);
	\end{tikzpicture}}
       \label{chp7_Appendix_fig:method_midpoint-dendrograma}
	\end{subfigure}
	\begin{subfigure}[t]{0.2\linewidth}
		\centering        
        \includegraphics[width=\linewidth, clip, trim={2.5cm 0cm 2.5cm 0cm}]{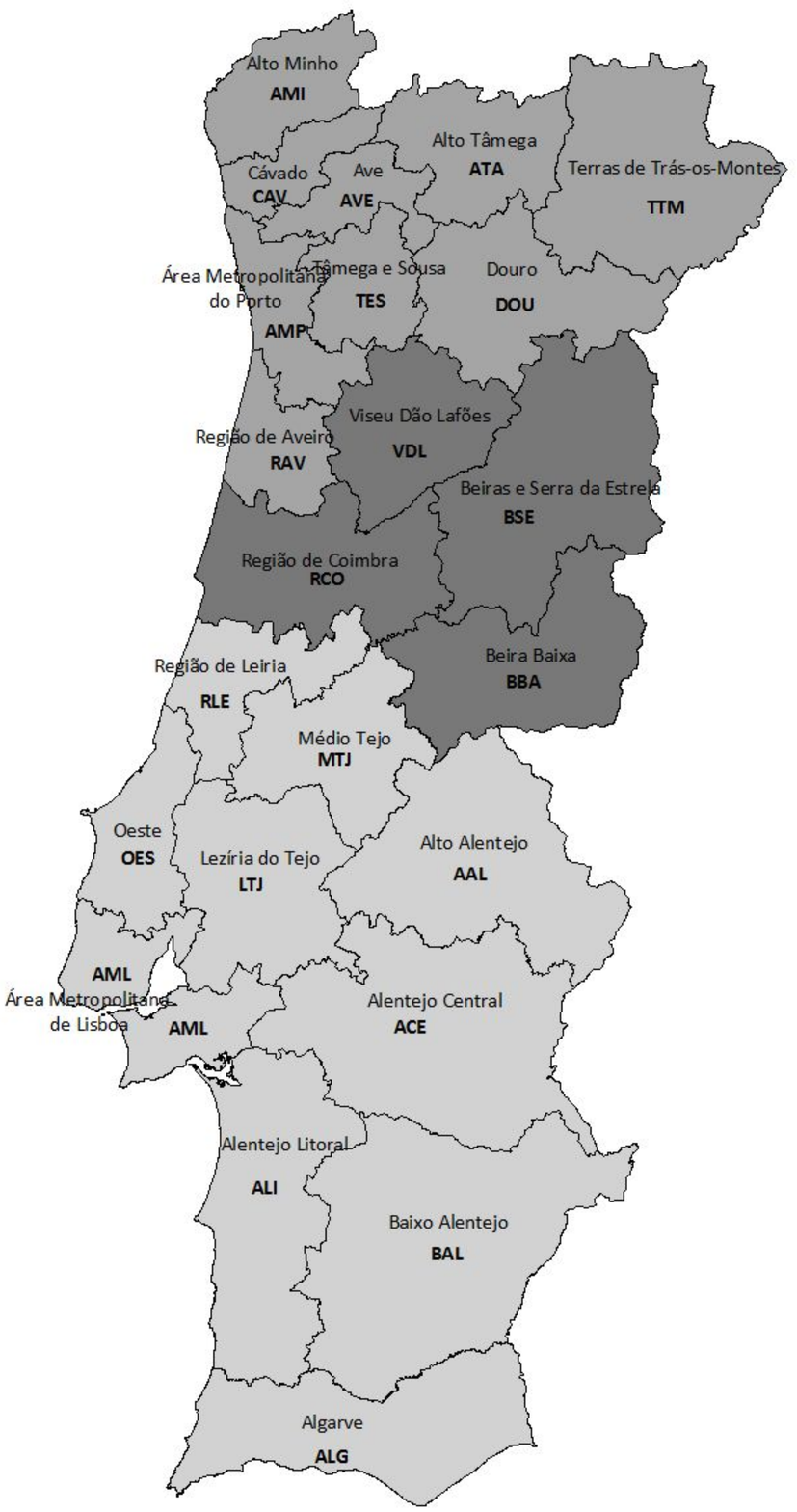}
        \label{chp7_Appendix_fig:method_midpoint-map2}
    \end{subfigure}	
\caption{On the left of the sketch for both methods is represented the final weighted network with super-vertices, followed on the right by the geographical representation of communities at level 1 (end of first \textit{pass}) of the LA, the Dendrogram for the Louvain's community detection process and, the geographical representation of communities at level 2 (end of second \textit{pass}) of the LA.}
  \label{chp7_Appendix_fig:method_midpoint}
\end{figure}

\end{document}